\documentclass[sigconf]{acmart}
\AtBeginDocument{%
  }

\setcopyright{acmlicensed}

\usepackage{listings}
\usepackage{enumitem}
\usepackage{amsfonts}
\usepackage{subcaption}
\usepackage{colortbl}
\usepackage{fancybox}
\usepackage{multirow}
\usepackage{xspace}
\usepackage{soul}
\usepackage{balance}

\newsavebox{\boxKbox}

\newenvironment{boxK}
{%
  \par\medskip\noindent%
  \setlength{\fboxsep}{10pt}
  \setlength{\fboxrule}{0.6pt}
  \begin{lrbox}{\boxKbox}%
    \begin{minipage}{\dimexpr\columnwidth-2\fboxsep-2\fboxrule\relax}%
      \small%
}
{%
    \end{minipage}%
  \end{lrbox}%
  \fcolorbox{black!25}{black!5}{\usebox{\boxKbox}}%
  \par\medskip%
}

\lstdefinestyle{javastyle}{
    language=Java,
    basicstyle=\ttfamily\small,
    keywordstyle=\color{blue}\bfseries,
    commentstyle=\color{gray},
    stringstyle=\color{red},
    numbers=left,
    numberstyle=\tiny,
    stepnumber=1,
    breaklines=true,
    frame=single
}

\newcommand{\ie}{\textit{i.e.,}\xspace}
\newcommand{\eg}{\textit{e.g.,}\xspace}

\newcommand{\etal}{et al.\xspace}

\newcommand\revision[1]{\textcolor{black}{#1}}
\newcommand\mrtwo[1]{\textcolor{black}{#1}}

\newcommand{\equref}[1]{Eq.~\ref{#1}\xspace}
\newcommand{\secref}[1]{Sec.~\ref{#1}\xspace}
\newcommand{\figref}[1]{Fig.~\ref{#1}\xspace}
\newcommand{\tabref}[1]{Table~\ref{#1}\xspace}

\newcommand{\circled}[1]{\textcircled{\small #1}}

\newcommand{\llm}{\textit{LLM}\xspace}
\newcommand{\llms}{\textit{LLMs}\xspace}
\newcommand{\llmsforcode}{\revision{LLM4Code}\xspace}
\newcommand{\llmsforcodes}{\revision{LLMs4Code}\xspace}
\newcommand{\sect}{SECT\xspace}
\newcommand{\rulerename}{\textit{Rule 1}\xspace}
\newcommand{\rulemdc}{\textit{Rule 13}\xspace}

\newcommand{\docode}{\textit{do$_{code}$}\xspace}
\newcommand{\scm}{\textit{SCM}\xspace}
\newcommand{\scms}{\textit{SCMs}\xspace}
\newcommand{\ate}{\textit{ATE}\xspace}
\newcommand{\ates}{\textit{ATEs}\xspace}

\newcommand{\loss}{\textit{LOSS}\xspace}
\newcommand{\mink}{\textit{MIN\_K}\xspace}
\newcommand{\zlib}{\textit{ZLIB}\xspace}

\newcommand{\codegen}{\textit{CodeGen}\xspace}
\newcommand{\codegpt}{\textit{CodeGPT}\xspace}
\newcommand{\starcoder}{\textit{starcoder2-3b}\xspace}
\newcommand{\starcoderseven}{\textit{starcoder2-7b}\xspace}
\newcommand{\codellama}{\textit{CodeLlama-7b}\xspace}
\newcommand{\deepseek}{\textit{deepseek-coder-1.3b}\xspace}
\newcommand{\mellum}{\textit{Mellum-4b}\xspace}
\newcommand{\stablecode}{\textit{stable-code-3b}\xspace}

\copyrightyear{2026}
\acmYear{2026}
\setcopyright{cc}
\setcctype{by}
\acmConference[ICSE '26]{2026 IEEE/ACM 48th International Conference on Software Engineering}{April 12--18, 2026}{Rio de Janeiro, Brazil}
\acmBooktitle{2026 IEEE/ACM 48th International Conference on Software Engineering (ICSE '26), April 12--18, 2026, Rio de Janeiro, Brazil}
\acmPrice{}
\acmDOI{10.1145/3744916.3787843}
\acmISBN{979-8-4007-2025-3/2026/04}

\begin{document}

\title{How Do Semantically Equivalent Code Transformations Impact Membership Inference on LLMs for Code?}

\author{Hua Yang}
\authornote{Both authors contributed equally to this research.}
\email{hyang45@ncsu.edu}
\orcid{0009-0000-1091-0495}
\affiliation{%
  \institution{North Carolina State University}
  \city{Raleigh}
  \state{North Carolina}
  \country{USA}
}

\author{Alejandro Velasco}
\authornotemark[1]
\email{svelascodimate@wm.edu}
\orcid{0000-0002-4829-1017}
\affiliation{%
  \institution{William \& Mary}
  \city{Williamsburg}
  \state{Virginia}
  \country{USA}
}

\author{Thanh Le-Cong}
\email{thanhlc@ieee.org}
\orcid{0000-0002-9566-324X}
\affiliation{%
  \institution{Singapore University of Technology and Design}
  \city{Singapore}
  \country{Singapore}
}

\author{Md Nazmul Haque}
\email{mhaque4@ncsu.edu}
\orcid{0000-0002-5120-3030}
\affiliation{%
  \institution{North Carolina State University}
  \city{Raleigh}
  \state{North Carolina}
  \country{USA}
}

\author{Bowen Xu}
\email{bxu22@ncsu.edu}
\orcid{0000-0002-1006-8493}
\affiliation{%
  \institution{North Carolina State University}
  \city{Raleigh}
  \state{North Carolina}
  \country{USA}
}
\authornote{Corresponding author.
}

\author{Denys Poshyvanyk}
\email{dposhyvanyk@wm.edu}
\orcid{0000-0002-5626-7586}
\affiliation{%
  \institution{William \& Mary}
  \city{Williamsburg}
  \state{Virginia}
  \country{USA}
}

\renewcommand{\shortauthors}{Yang, Velasco, et al.}

\begin{abstract}
\label{sec:abstract}
The success of large language models for code relies on vast amounts of code data, including public open-source repositories, such as GitHub, and private, confidential code from companies. This raises concerns about intellectual property compliance and the potential unauthorized use of license-restricted code. While membership inference (MI) techniques have been proposed to detect such unauthorized usage, their effectiveness can be undermined by semantically equivalent code transformation techniques, which modify code syntax while preserving semantic.

In this work, we systematically investigate whether semantically equivalent code transformation rules might be leveraged to evade MI detection. The results reveal that model accuracy drops by only 1.5\% in the worst case for each rule, demonstrating that transformed datasets can effectively serve as substitutes for fine-tuning. Additionally, we find that one of the rules (RenameVariable) reduces MI success by 10.19\%, highlighting its potential to obscure the presence of restricted code. To validate these findings, we conduct a causal analysis confirming that variable renaming has the strongest causal effect in disrupting MI detection. Notably, we find that combining multiple transformations does not further reduce MI effectiveness. Our results expose a critical loophole in license compliance enforcement for training large language models for code, showing that MI detection can be substantially weakened by transformation-based obfuscation techniques.
\end{abstract}



\maketitle

\section{Introduction}
\label{sec:introduction}


Recent advances in \textit{ large language models for code (\llmsforcode)} have significantly improved automated code completion, generation, and understanding~\cite{li2023starcoder, deng2024can, nijkamp2023codegen}. These models employ complex architectures and are trained on extensive code corpora, often sourced from open-source repositories such as GitHub~\cite{gao2020pile, husain2019codesearchnet, allamanis2013mining}. However, their dependence on large-scale data raises critical concerns about intellectual property compliance and the potential unauthorized use of license-restricted code.
The debate on developers' rights to be informed about using their code for training language models remains ongoing~\cite{githubissuelicense}.
To address this issue, dataset curators often incorporate license restrictions when collecting code from open-source repositories. For example, The Stack~\cite{Kocetkov2022TheStack}, one of the largest publicly available code datasets, is comprised of more than 6TB of source code covering 358 programming languages and prioritizes permissively licensed code during data collection. However, enforcing license compliance remains a challenge, as malicious LLM developers can still extract and utilize license-restricted code for training, bypassing ethical and legal constraints.
In this work, we consider a malicious scenario in which an LLM developer intends to build a model for code by illegally using restricted code without being detected by Membership Inference (MI) techniques. Our aim is to assess the risk of such a threat.

To mitigate such risks, recent studies have explored MI techniques as a detection mechanism~\cite{wan2024does, yang2024gotcha, li2025investigating, haque2025quantization}. MI techniques aim to determine whether specific code snippets were included in an \llmsforcode's training data. Previous research has shown that MI can effectively identify unauthorized use of license-restricted code, highlighting its potential to audit and enforce compliance in large-scale model training~\cite{yang2024gotcha}.



Semantically Equivalent Code Transformation (\revision{\sect}) techniques pose a potential risk for malicious purposes due to their ability to obfuscate original code to appear syntactically different (while preserving the same functionality and semantics). The rationale behind this concern is threefold. \revision{First, recent empirical studies~\cite{orvalho2025large} have shown that semantics-preserving transformations—such as reordering statements or renaming variables—do not significantly degrade the performance of \llmsforcode, suggesting that training on a dataset with such transformations applied will yield similar model performance. This observation is further supported by our own experiments, which show a minimal change in accuracy on the code completion task.} This means that malicious LLM developers relying on transformed restricted code face little risk of a significant performance drop. Second, \revision{\sect} can influence LLM memorization, potentially impacting the effectiveness of MI techniques. Recent findings support this claim, showing that LLMs exhibit a strong memorization of specific patterns and training data samples~\cite{yang2024unveiling, karmakar2022codex}. Third, a fundamental limitation of standard MI \revision{techniques} is their reliance on exact match detection~\cite{duan2024membership}. In other words, an attacker can potentially evade MI detection by applying \textbf{\revision{\sect}}—modifying code syntax while retaining its original functionality and semantics. These transformations can obfuscate the presence of license-restricted code in the training data, complicating the detection of unauthorized data usage. Despite the potential impact of \revision{\sect} on MI, its effect has yet to be empirically validated.
Hence, we propose our central research question (RQ):

\begin{boxK}
\textit{How do semantically equivalent program transformation techniques impact membership inference?}
\end{boxK}



To address this RQ, we first collected 23 \revision{\sect} rules applicable to Java to generate 23 new semantically equivalent Java datasets. \revision{We then fine-tuned \mrtwo{eight} \llmsforcode, namely \codegpt~\cite{lu2021codexglue}, \codegen~\cite{nijkamp2023codegen}, \deepseek~\cite{guo2024deepseek}, \stablecode~\cite{stable-code-3b}, \mrtwo{\mellum~\cite{Mellum-4b-base}}, \starcoder, \mrtwo{\starcoderseven~\cite{lozhkov2024starcoder}},  and \codellama~\cite{roziere2023code} on these 25 datasets (including the original one and the one that applied all rules). The results demonstrate that models fine-tuned on semantically equivalent datasets exhibit highly consistent performance: 135 out of 138 models experienced no more than a 1\% drop in accuracy, with the maximum observed degradation being only 1.5\%. These findings confirm the feasibility of using such transformed datasets as substitutes for fine-tuning.}

\revision{Next, we evaluated the impact of \revision{\sect} on MI performance. We found that fine-tuning \deepseek with a dataset transformed by \rulerename resulted in a 10.19\% decrease in MI detection, while the task performance dropped by only 0.63\%. This substantial disparity raises concerns about the potential of \revision{\sect} to circumvent license enforcement mechanisms.} We then examined whether these rules exhibit cumulative effects on MI performance and found that they do not, which is reassuring. Finally, we conducted a causal analysis using a Structural Causal Model (\scm) to validate our findings, further strengthening the credibility of our conclusions. To summarize, the key contributions of this paper are as follows:


\begin{itemize}
\item \textbf{Impact of \revision{\sect} Rules on MI Performance:} We are the first to investigate the risks of using \revision{\sect} to bypass MI detection, which allows training \llmsforcode without complying with licensing requirements. Our results demonstrate that \rulerename has the most significant impact on MI performance, in which \revision{10.19\% more} of code remains undetected by MI, indicating a high risk of licensing breaches.

\item \textbf{Causal Analysis of MI Degradation:} Our causal analysis confirms that \rulerename \textit{RenameVariable} has the strongest effect in reducing MI success, directly disrupting model memorization. While other transformations (\eg \textit{Rule 7}, \textit{Rule 8}, and \rulemdc) have moderate effects, combining all transformations does not provide an additional reduction in MI effectiveness. Refutation tests validate these causal findings, supporting the reliability of the estimated causal effects.

\item \textbf{Replication Package:} To support further research on studying the effect of \sect on MI and facilitating the development of effective defenses, we publicly release our transformation rules, causal analysis tools, source code and data used in the experiment \footnote{\url{https://github.com/ncsu-softmax/mia_codetrans_causal}}.
\end{itemize}

\section{Background}
\label{sec:background}

\revision{We adopt the perspective of a defender who leverages MI techniques to audit model behavior and detect potential misuse of protected training data. Our focus is on the fine-tuning stage, where datasets are typically smaller, domain specific, and more likely to contain sensitive or proprietary content~\cite{fu2024membership, yu2022differentially, yang2025understanding}, thereby presenting a higher risk of adversarial misuse.}

\revision{A well-defined threat model, which outlines the attacker’s knowledge, assumptions, and capabilities, is essential to any rigorous security analysis.} In this section, we first outline the threat model. We then review existing MI techniques, describe the memorization behavior of \llmsforcodes in code generation, and introduce causal interpretability as a framework to assess the effectiveness of \sect in lowering MI performance.




\subsection{Threat Model}

    \revision{We consider a threat model where the adversary's objective is to circumvent licensing restrictions on protected code by using \sect before using it to fine-tune a language model for code completion. From the defender’s perspective, we employ MI techniques to audit model behavior and detect whether protected training data may have been improperly used.} 
    \revision{In this black-box setting, the defender can query the model and observe its output and confidence scores (\eg logits or loss values)~\cite{fu2024membership, duan2024membership, mattern2023membership}, but has no access to the internal model parameters.}

This setting reflects realistic deployment scenarios both commercial and open source \llmsforcodes. For example, commercial models such as OpenAI's GPT-4~\cite{achiam2023gpt} expose only limited information via APIs, while open-source models such as StarCoder~\cite{li2023starcoder} and CodeLlama~\cite{roziere2023code} are often evaluated under similar query constraints to simulate practical audit conditions.

\begin{figure}[t]
    \centering
    \includegraphics[width=\columnwidth]{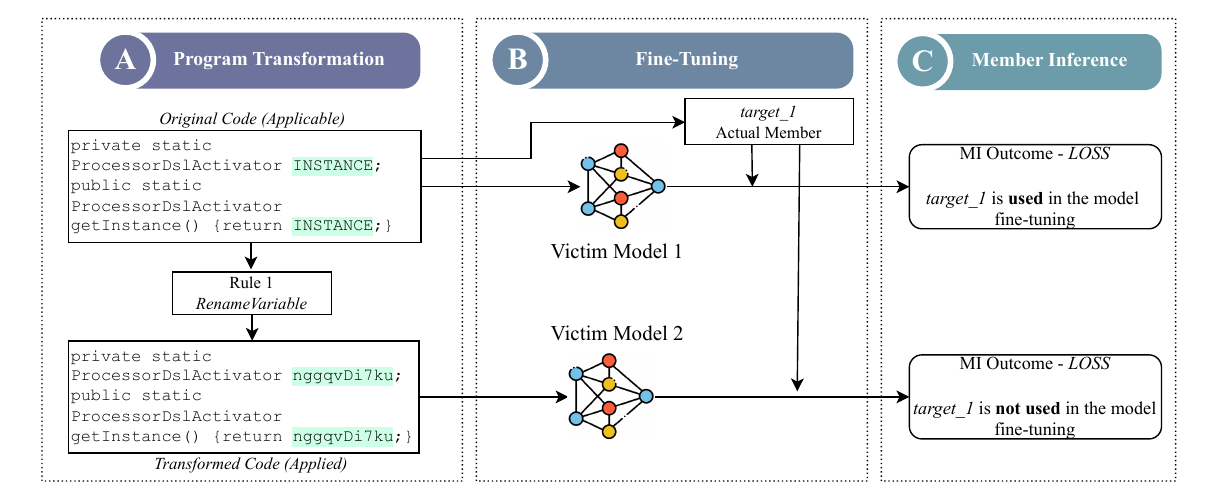}
    \caption{\revision{An example where a membership inference method successfully predicted membership before transformation but failed after transformation}.}
    \Description{The figure shows an example of membership inference before and after SECT. It illustrates that variable renaming preserves \llmsforcodes performance but causes the membership inference method to incorrectly classify the original member code as a non-member.}

    \label{fig:example}
\end{figure}
We provide an actual example in~\figref{fig:example}. The original code is used to fine-tune \codegpt (victim model 1) and is accurately detected by \loss (MI method; see Section~\ref{sec:methodology}) as a member, which means that it is correctly identified as being used in the fine-tuning process. However, after applying \textit{RenameVariable} rule to the original code, the transformed version is used to fine-tune another \codegpt (victim model 2). Despite victim model 2 achieving similar performance to victim model 1, the \loss mistakenly classifies the original code as a non-member. As a result, the malicious developer bypasses the licensing restrictions on the original code.


\subsection{Membership Inference Techniques}


\revision{MI is the task of determining whether a specific data sample was part of the training dataset. While MI has been studied primarily in the context of privacy attacks, it can also serve as a valuable analytical tool to detect whether sensitive or unauthorized data has been used during training. MI is effective for LLMs trained on proprietary or sensitive codebases, where successful inference can reveal the inclusion of confidential or security-critical information~\cite{carlini2021extracting, yang2024gotcha}. This enables researchers and practitioners to better understand which models may have been trained on inappropriate or license-restricted content.}


Current MI approaches can be broadly classified into two categories. The first category, direct analysis, infers membership by examining statistical patterns in the model’s predicted outputs. Initially introduced by Shokri \etal~\cite{shokri2017membership}, these methods exploit differences in model behavior between training and unseen samples, often relying on confidence scores, loss values, or other statistical measures. Notable examples include \loss, \zlib, and \mink. The second category, classifier-based inference, extends this idea by using the predicted output of the model as input to a trained binary classifier to distinguish between member and nonmember data, ultimately assigning a membership label (1 for member, 0 for nonmember). Notable classifier-based methods include GOTCHA~\cite{yang2024gotcha} and CODEMI~\cite{wan2024does}.

\subsection{Causal Interpretability}
Causal interpretability, first introduced in \docode \cite{docode}, is a post hoc interpretability framework leveraging Pearl's ladder of causation \cite{Pearl2018Causality} to establish cause-and-effect relationships in the behavior of the model for SE tasks, distinguishing genuine causal effects from spurious correlations. In the context of security for \llms, we contextualize \docode to analyze how \revision{\sect} influences a model’s ability to retain or obscure training data. Specifically, we apply statistical and causal inference techniques to measure the \textit{causal effect} of \revision{\sect} on the MI outcomes. By formulating causal hypotheses using \scm, we estimate the impact of transformations while controlling for confounders, ensuring a rigorous evaluation of whether observed effects reflect genuine changes in MI effectiveness or superficial variations. Following \docode, we \circled{1} \textbf{model} the inference problem using causal assumptions encoded in graph representations, \circled{2} \textbf{identify} the causal estimand, \circled{3} \textbf{estimate} causal effects using probabilistic and machine learning methods, and \circled{4} \textbf{validate} causal estimates through sensitivity analyzes, thereby ensuring the robustness of computed causal effects.

\section{Methodology}
\label{sec:methodology}

\begin{figure}[ht]
		\centering
  \includegraphics[width=.48\textwidth]{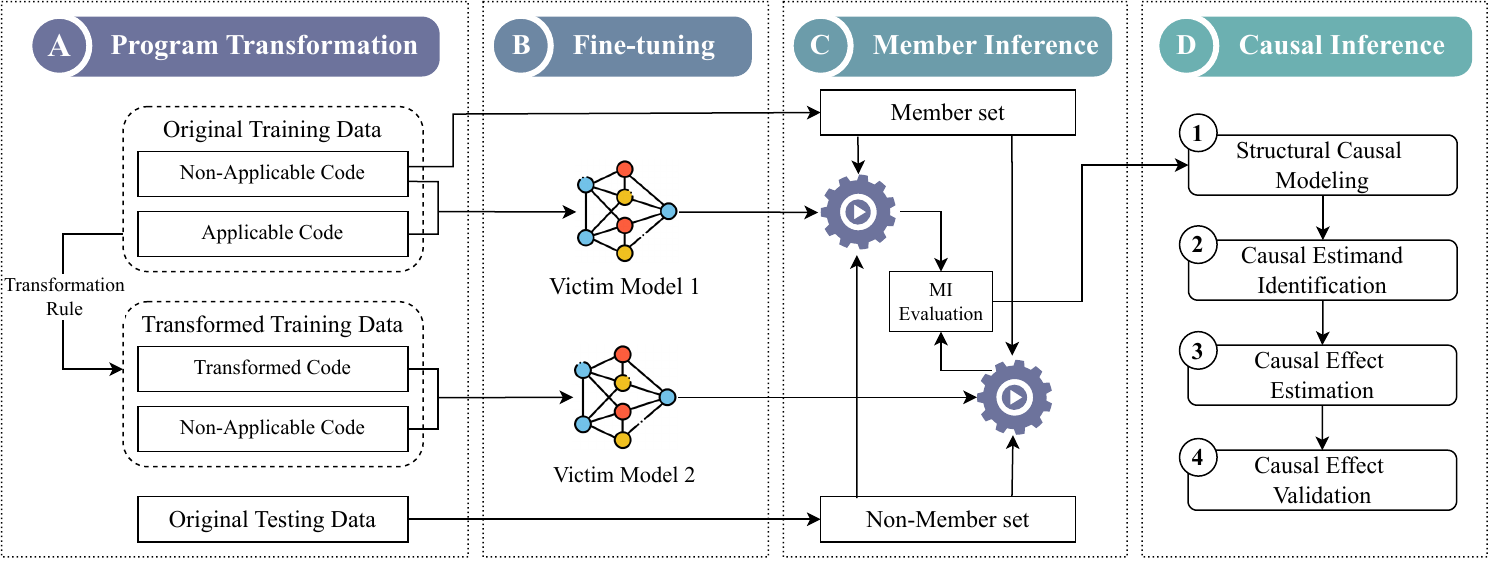}
		\caption{Experimental workflow of our study. We begin with program transformation and fine-tuning, followed by membership inference, and conclude with a causal analysis of how code transformations influence MI effectiveness.}
        \Description{The figure presents the experimental workflow, including code transformation, model fine-tuning, membership inference evaluation, and causal analysis. Arrows indicate the data flow and dependencies between each stage.}
    \label{fig:pipeline}
\end{figure}

In this section, we provide an overview of our methodology. Our experimental workflow is illustrated in~\figref{fig:pipeline}. Our procedure begins with program transformation, in which we apply the \revision{\sect} rules to modify the original code. Next, we fine-tune the victim models using either the original or transformed code. We then evaluate MI to assess the impact of these transformations on the success rate. Finally, we performed a causal analysis to further investigate our findings, providing insight into \textit{how} and \textit{why} certain transformations succeed in evading MI detection.


\subsection{Semantically Equivalent Code Transformation}



This study investigates whether code transformations pose the risk of enabling unauthorized use of license-restricted code. For this to hold, two conditions must be met: (1) the model’s performance after training should remain comparable to its original performance, and (2) the MI success rate should be reduced. If we can generate code that is functionally identical to the target code while appearing distinct, it should be possible to satisfy these conditions.

To achieve this, we synthesize semantically equivalent variants of the original code for training. Since determining program equivalence is undecidable~\cite{kozen1977rice}, we employ hand-crafted transformation rules to ensure semantic preservation. 
In total, we collected 23 transformation rules, carefully designed and validated by previous work~\cite{letowards, zhang2023challenging, yu2022data, rabin2021generalizability}. These rules cover multiple levels of granularity, including naming, expression, and statement, ensuring a diverse set of transformations. A comprehensive list of these transformation rules is provided in~\tabref{tab:code_transformations}. \revision{We have provided a concrete example for each rule in our replication package.} Finally, we apply these rules to the GitHub Java corpus to generate semantically equivalent datasets.

\begin{table}[htbp]
\centering
\caption{Code transformations at different levels.}
\Description{The table lists 23 semantically equivalent code transformation rules for java, describing their functionality and the program granularity level at which each rule operates.}
\label{tab:code_transformations}
\vspace{-1em}
\scalebox{.533}{
\setlength{\tabcolsep}{4pt} 

\begin{tabular}{cllllll}
\multicolumn{1}{l}{\textbf{ID}} &
   &
  \textbf{Transformation Rule} &
   &
  \textbf{Description} &
   &
  \textbf{Level} \\ \hline
\textit{1} &
   &
  \textit{RenameVariable} &
   &
  Renames a variable while preserving functionality. &
   &
  Naming \\
\rowcolor[HTML]{EFEFEF} 
\textit{2} &
   &
  \textit{For2While} &
   &
  Converts a \textbackslash{}texttt\{\textbackslash{}small 'for'\} loop into a \textbackslash{}texttt\{\textbackslash{}small 'while'\} loop. &
   &
  Statement \\
\textit{3} &
   &
  \textit{While2For} &
   &
  Converts a \textbackslash{}texttt\{\textbackslash{}small 'while'\} loop into a \textbackslash{}texttt\{\textbackslash{}small 'for'\} loop. &
   &
  Statement \\
\rowcolor[HTML]{EFEFEF} 
\textit{4} &
   &
  \textit{Do2While} &
   &
  \begin{tabular}[c]{@{}l@{}}Transforms a \textbackslash{}texttt\{\textbackslash{}small 'do-while'\} loop into a \\ \textbackslash{}texttt\{\textbackslash{}small 'while'\} loop.\end{tabular} &
   &
  Statement \\
\textit{5} &
   &
  \textit{IfElseIf2IfElse} &
   &
  \begin{tabular}[c]{@{}l@{}}Refactors an \textbackslash{}texttt\{\textbackslash{}small 'if-else-if'\} chain into a nested \\ \textbackslash{}texttt\{\textbackslash{}small 'if-else'\}.\end{tabular} &
   &
  Statement \\
\rowcolor[HTML]{EFEFEF} 
\textit{6} &
   &
  \textit{IfElse2IfElseIf} &
   &
  \begin{tabular}[c]{@{}l@{}}Converts an \textbackslash{}texttt\{\textbackslash{}small 'if-else'\} structure into an \\ \textbackslash{}texttt\{\textbackslash{}small 'if-else-if'\} chain.\end{tabular} &
   &
  Statement \\
\textit{7} &
   &
  \textit{Switch2If} &
   &
  \begin{tabular}[c]{@{}l@{}}Replaces a \textbackslash{}texttt\{\textbackslash{}small 'switch'\} statement with equivalent \\ \textbackslash{}texttt\{\textbackslash{}small 'if-else'\} logic.\end{tabular} &
   &
  Statement \\
\rowcolor[HTML]{EFEFEF} 
\textit{8} &
   &
  \textit{Unary2Add} &
   &
  Replaces a unary operation with an equivalent addition. &
   &
  Expression \\
\textit{9} &
   &
  \textit{Add2Equal} &
   &
  \begin{tabular}[c]{@{}l@{}}Converts an addition assignment (\textbackslash{}texttt\{\textbackslash{}small '+='\}) into an equality \\ assignment (\textbackslash{}texttt\{\textbackslash{}small '=='\}).\end{tabular} &
   &
  Expression \\
\rowcolor[HTML]{EFEFEF} 
\textit{10} &
   &
  \textit{DivideVarDecl} &
   &
  Splits multiple variable declarations into separate statements. &
   &
  Expression \\
\textit{11} &
   &
  \textit{MergeVarDecl} &
   &
  Combines multiple variable declarations into a single statement. &
   &
  Expression \\
\rowcolor[HTML]{EFEFEF} 
\textit{12} &
   &
  \textit{SwapStatement} &
   &
  Swaps the order of two statements without data-flow dependencies. &
   &
  Statement \\
\textit{13} &
   &
  \textit{ModifyConstant} &
   &
  Replaces a constant value with an equivalent mathematical expression. &
   &
  Expression \\
\rowcolor[HTML]{EFEFEF} 
\textit{14} &
   &
  \textit{ReverseIf} &
   &
  \begin{tabular}[c]{@{}l@{}}Reverses the condition in an \textbackslash{}texttt\{\textbackslash{}small 'if'\} statement and swaps \\ its branches.\end{tabular} &
   &
  Statement \\
\textit{15} &
   &
  \textit{If2CondExp} &
   &
  \begin{tabular}[c]{@{}l@{}}Converts an \textbackslash{}texttt\{\textbackslash{}small 'if'\} statement into a conditional \\ (\textbackslash{}texttt\{\textbackslash{}small '?:'\}) expression.\end{tabular} &
   &
  Statement \\
\rowcolor[HTML]{EFEFEF} 
\textit{16} &
   &
  \textit{ConfExp2If} &
   &
  \begin{tabular}[c]{@{}l@{}}Converts a conditional (\textbackslash{}texttt\{\textbackslash{}small '?:'\}) expression into \\ an \textbackslash{}texttt\{\textbackslash{}small 'if'\} statement.\end{tabular} &
   &
  Statement \\
\textit{17} &
   &
  \textit{InfixDividing} &
   &
  Splits an infix expression into smaller sub-expressions. &
   &
  Expression \\
\rowcolor[HTML]{EFEFEF} 
\textit{18} &
   &
  \textit{DividePrePostFix} &
   &
  Separates pre/postfix expressions into explicit statements. &
   &
  Expression \\
\textit{19} &
   &
  \textit{DividingComposedIf} &
   &
  \begin{tabular}[c]{@{}l@{}}Breaks down a complex \textbackslash{}texttt\{\textbackslash{}small 'if'\} condition into multiple \\ conditions.\end{tabular} &
   &
  Statement \\
\rowcolor[HTML]{EFEFEF} 
\textit{20} &
   &
  \textit{LoopIfContinue2Else} &
   &
  \begin{tabular}[c]{@{}l@{}}Converts a \textbackslash{}texttt\{\textbackslash{}small 'continue'\} inside a loop into an \\ \textbackslash{}texttt\{\textbackslash{}small 'else'\} branch.\end{tabular} &
   &
  Statement \\
\textit{21} &
   &
  \textit{SwitchEqualExp} &
   &
  \begin{tabular}[c]{@{}l@{}}Modifies a \textbackslash{}texttt\{\textbackslash{}small 'switch'\} expression based on equality \\ comparisons.\end{tabular} &
   &
  Expression \\
\rowcolor[HTML]{EFEFEF} 
\textit{22} &
   &
  \textit{SwitchStringEqual} &
   &
  \begin{tabular}[c]{@{}l@{}}Converts a \textbackslash{}texttt\{\textbackslash{}small 'switch'\} on strings into equivalent \\ \textbackslash{}texttt\{\textbackslash{}small 'if-else'\} conditions.\end{tabular} &
   &
  Expression \\
\textit{23} &
   &
  \textit{SwitchRelation} &
   &
  \begin{tabular}[c]{@{}l@{}}Transforms a \textbackslash{}texttt\{\textbackslash{}small 'switch'\} statement based on \\ relational operators.\end{tabular} &
   &
  Expression \\ \hline
\end{tabular}
} 
\end{table}


\subsection{Selected Membership Inference Techniques}
\label{sec:mi}

In this study, the objective of MI is to determine whether a given code snippet \( x \) is part of the training dataset of an LLM, \( M \). This involves computing a membership score \( f(x, M) \), which is then thresholded to infer the membership status of \( x \). We consider three MI techniques, each employing a distinct scoring function \( f(x, M) \):

\begin{itemize}[wide=0pt]
    \item \textbf{\loss} (\cite{yeom2018privacy}): This \revision{method} assumes that training samples typically exhibit lower loss values compared to nonmembers. The membership score is computed using the model’s loss function:

    \begin{equation} 
        \label{eq:loss}
        f_{\text{loss}}(x, M) = \mathcal{L}(x; M)
    \end{equation}

    where \( \mathcal{L}(x; M) \) is the average token-level negative log-likelihood (e.g., cross-entropy loss) for sample \( x \).
    
    \revision{\textbf{\mink}~\cite{shi2023detecting}: This method is based on the observation that non-member examples tend to contain more low-likelihood tokens when evaluated by the target model \( M \). Instead of averaging over all tokens, it selects the \(k\%\) tokens with the lowest predicted probabilities (i.e., highest negative log-likelihoods). The membership score is then computed as:}
\revision{
    \begin{equation}
    \label{eq:min_k}
    f_{\text{min-}k}(x; M) = \frac{1}{|\text{min-}k(x)|} \sum_{x_i \in \text{min-}k(x)} -\log p(x_i \mid x_1, \ldots, x_{i-1}),
    \end{equation}
    }
    \revision{where $p(x_i \mid x_1, \ldots, x_{i-1})$ denotes the model $M$ assigned probability of token $x_i$ given its left-side context, and \( \text{min-}k(x) \) denotes the set of \(k\%\) tokens in \( x \) with the lowest model-assigned probabilities.}
    
    \revision{\textbf{\zlib}~\cite{carlini2021extracting}: This method calibrates the model-assigned loss in \loss using the Zlib-compressed size of the input sequence, which serves as a proxy for input complexity. The membership score is defined as:}
    \revision{
    \begin{equation}
    \label{eq:zlib}
    f_{\text{zlib}}(x; M) = \frac{\mathcal{L}(x; M)}{\text{zlib}(x)},
    \end{equation}
    }
\revision{where \( \textit{zlib}(x) \) is the length in bytes of the zlib compressed sample.} 

\end{itemize}



\subsection{Structural Causal Modeling}
\label{sec:scm}

\begin{figure}[ht]
		\centering
  \includegraphics[width=0.45\textwidth]{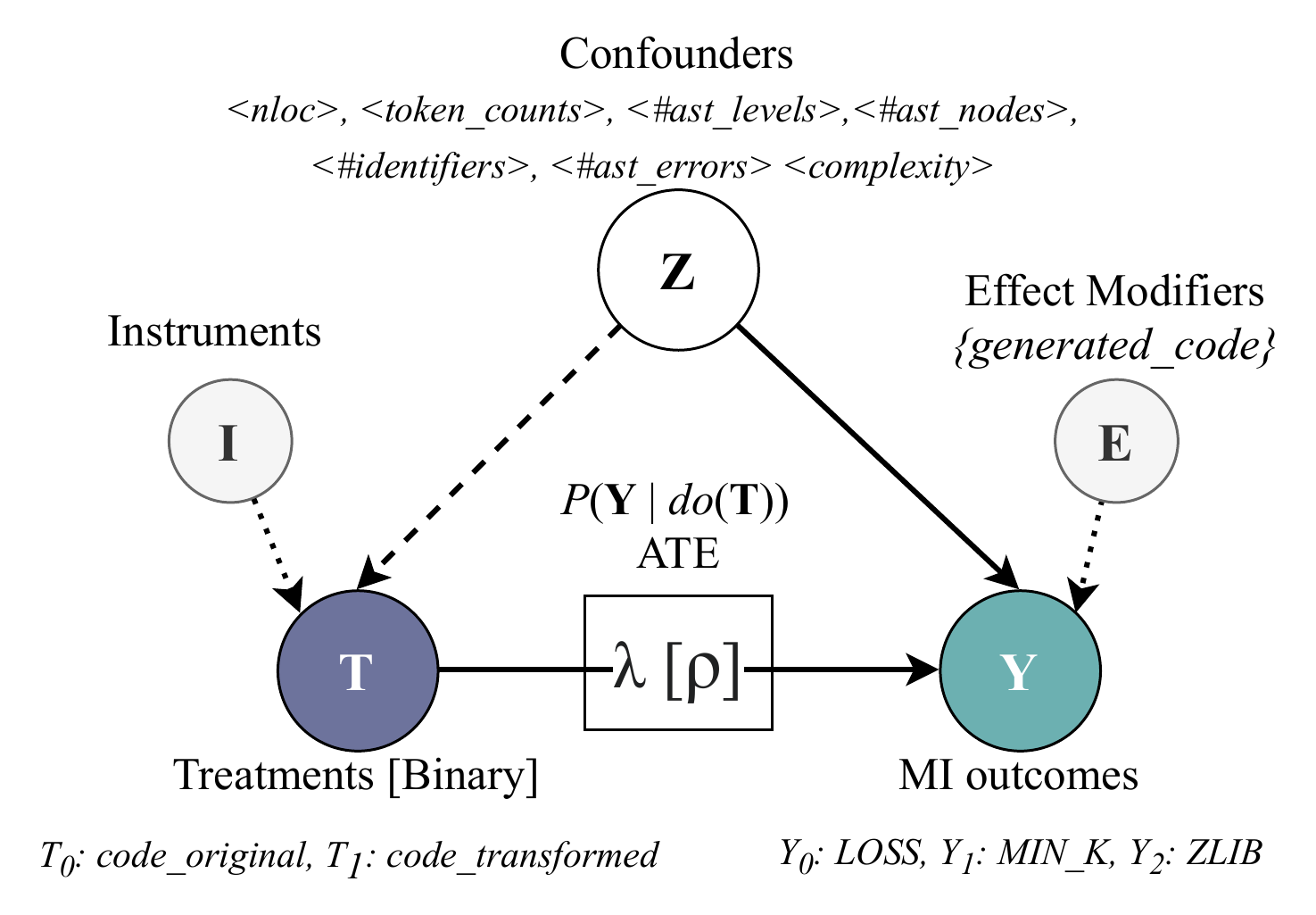}
  \vspace{-3mm}
		\caption{\revision{Structural Causal Model for MI.}}
        \Description{The figure depicts a structural causal model illustrating the relationship between code transformations, confounding variables, and membership inference outcomes.}
    \label{fig:structural_causal_model}
\vspace{-1.5em}
\end{figure}

To investigate the causal relationship between the code transformations in \tabref{tab:code_transformations} and the effectiveness of MI, we follow the Causal Inference steps illustrated in \figref{fig:pipeline}. First, we propose \scm. \scms allow us to explicitly encode our causal hypothesis and analyze causal relationships among variables. By clearly defining treatments, outcomes, and confounders, \scms facilitate precise estimation of causal effects while mitigating biases inherent in observational data. In particular, \figref{fig:structural_causal_model} illustrates the \scm proposed for our study, encoding our primary causal hypothesis: \textbf{\textit{altering the code used for fine-tuning through \revision{\sect} rules impacts MI effectiveness}}.

In the proposed \scm, binary treatments ($T$), specifically the original code ($T_0$) and semantically equivalent transformed code ($T_1$), directly influence MI potential outcomes, measured by three metrics: $Y_0$ (\loss in \equref{eq:loss}), \revision{$Y_1$ (\mink in \equref{eq:min_k}) and $Y_2$ (\zlib in \equref{eq:zlib})}. The causal effect between treatments and outcomes is influenced by confounders ($Z$), corresponding to variables extracted from the source code, such as \textit{<nloc>} (\ie number of lines of code), \textit{<token\_counts>} (\ie number of tokens), \textit{<\#ast\_levels>} (\ie height of AST), \textit{<\#ast\_nodes>}, \textit{<\#identifiers>}, \textit{<\#ast\_errors>} (\ie number of AST parsing errors), and \textit{<code\_complexity>}. In addition, variables present in the generated code serve as effect modifiers, influencing the magnitude or direction of the causal relationship between treatments and outcomes. Instrumental variables, commonly used to address confounding bias by isolating variations independent of confounders, were not considered in this study. 

\textbf{Computing \ates}: \textit{Average Treatment Effect} (\ate) denoted by ($\lambda [\rho]$ in \figref{fig:structural_causal_model}), quantifies the expected difference in MI outcomes between transformed and original code scenarios, averaged over the distribution of confounders and effect modifiers. Formally, \ate is computed using Pearl's \textit{do-operator} \cite{Pearl2018Causality} as $P(Y \mid do(T))$, representing an intervention scenario where treatments are set independently of confounders (\eg Binary Treatment). This computation isolates the causal effect of code transformations by eliminating biases due to confounders. To estimate \ates, a causal estimand is identified using graph-based criteria and the \textit{adjustment formula}, determining the appropriate subset of confounders $Z$ from \scm. Specifically, the identification of the causal estimand uses the \textit{ doWhy} library to systematically explore candidate causal estimands using criteria such as the backdoor criterion, frontdoor criterion, instrumental variables and mediation analysis~\cite{dowhy}.

\textbf{Measuring \ate's Robustness:} After computing the \ate of each transformation rule, we perform sensitivity analysis using multiple refutation techniques to validate the robustness of the obtained results. Specifically, we apply the following methods: ($R_1$) \textit{introducing a random common cause} by adding an independent variable to check if the effect remains stable, ensuring it is not due to spurious correlations. ($R_2$) applying \textit{placebo treatments} by replacing the actual treatment with a random one, where a near-zero effect confirms the validity of the original estimate. ($R_3$) adding \textit{unobserved confounders} correlated with treatments and outcomes to assess sensitivity, minimal effect change indicates robustness. And ($R_4$) performing \textit{data subset} validation by checking effect consistency across subsets, where stability reinforces generalizability.

\textbf{Interpreting \ates:} The \ate represents the expected difference in outcomes between treated and untreated groups, providing an estimate of causal impact. A positive \ate suggests that the applied transformation rule increases the effectiveness of MI, while a negative \ate indicates a reduction. The magnitude of the \ate reflects the strength of the effect, but its reliability depends on robustness checks and statistical significance.

\section{Study Design}
\label{sec:exp_settings}

To systematically assess the impact of \revision{\sect} on MI against fine-tuned \llmsforcodes, we design a comprehensive empirical study. Our study investigates the effect of individual and combined transformations on MI effectiveness and provides a causal interpretation of their impact. This section outlines our research questions (RQs), the models and datasets used, evaluation metrics, causal analysis methodology, and implementation details.


\subsection{Research Questions}
\begin{enumerate}[label=\textbf{RQ$_{\arabic*}$}, ref=\textbf{RQ$_{\arabic*}$}, wide=0pt, labelindent=5pt]\setlength{\itemsep}{0.2em}
      \item \label{rq:performance_1} {\textbf{[Impact of Individual Rules on MI]} How does each transformation rule impact membership inference?}
      \noindent\textbf{Motivation}: \revision{When performing \revision{\sect} on a code snippet, the transformed version should preserve the same functionality and semantics as the original code.} However, since each transformation rule modifies the code differently, their effects on the model may vary. For example, Rule~1 (\textit{RenameVariable}) replaces variable names with randomly generated strings, whereas Rule~2 (\textit{For2While}) modifies the control flow by converting \texttt{for} loops into \texttt{while} loops. \revision{Each transformation introduces a distinct type of modification, which potentially alters the training data's token-level likelihoods or the model's loss surface. However, it remains unclear whether such changes are sufficient to affect the outcome of MI techniques. Therefore, it is essential to investigate how different \sect rules impact MI performance.}

      \item \label{rq:performance_2} {\textbf{[Impact of Combined Rules on MI]} How do combined transformation rules impact membership inference?}
      \noindent\textbf{Motivation}: Applying more rules leads to a greater difference from the original code. Intuitively, this may make MI techniques less effective in distinguishing members from nonmembers. Moreover, considering that a transformation rule may not be applicable on a certain code, malicious LLM developers may wish to apply other rules. Hence, we aim to investigate if \textit{cumulative effects}  exist when multiple rules are applied.
      \item \label{rq:causality} {\textbf{[Interpret the Impact by Causality]} To what extent does each transformation rule impact the success of MI?}
      \noindent\textbf{Motivation}: To validate the results of \ref{rq:performance_1} and \ref{rq:performance_2} and assess the true impact of each transformation on the effectiveness of MI, we move beyond correlation analysis to \textit{causal inference}. Correlations may indicate relationships, but cannot confirm whether transformations genuinely reduce the MI success rate or if confounding factors influence the results.

\end{enumerate}



\subsection{Models}


In this study, we focus on fine-tuning LLMs for code, as fine-tuning has been extensively studied in prior research~\cite{yang2024gotcha, fu2024membership, wan2024does}. For the downstream task, we select code completion, a core feature in nearly all modern integrated development environments. Code completion suggests relevant code according to the current coding context. Analyzing token-level completion allows us to examine how these transformations impact fine-tuned model's performance and the success rate of membership inference.

We do not use models such as OpenAI Codex or ChatGPT, as their training data have not been publicly released, making it impossible to evaluate membership leakage~\cite{yang2024gotcha}. Instead, we select \revision{\codegpt~\cite{lu2021codexglue}, \codegen~\cite{nijkamp2023codegen}, \deepseek~\cite{guo2024deepseek}, \stablecode~\cite{stable-code-3b}, \mrtwo{\mellum~\cite{Mellum-4b-base}}, \starcoder, \mrtwo{\starcoderseven~\cite{lozhkov2024starcoder}},  and \codellama~\cite{roziere2023code} as our victim models.}
\codegpt is a transformer-based language model with the same architecture and training objective as GPT-2 \cite{radford2019language}. It is pre-trained on the CodeSearchNet dataset. 
\codegen uses an autoregressive transformer architecture with next-token prediction as its learning objective. 
\revision{
\deepseek is an open-source code language model trained on 2T tokens, with a composition of 87\% code and 13\% natural language in both English and Chinese. 
\mrtwo{\starcoder and \starcoderseven are decoder-only models trained on an impressive corpus of 17 programming languages.
\mellum is JetBrains' first open-source code language model with 4 billion parameters, specifically optimized for code completion tasks. The model adopts a LLaMA-style architecture and supports multiple programming languages.}
\stablecode is a 2.7B billion parameter decoder-only language model pre-trained on 1.3 trillion tokens of diverse textual and code datasets. 
\codellama is a the base 7B version of pretrained and fine-tuned generative text models. This model is designed for general code synthesis and understanding, supporting languages such as Python, C++, and Java.
}


\subsection{Datasets}
\label{sec:datasets}

We consider a widely used dataset from the CodeXGLUE benchmark: the GitHub Java Corpus, collected by Allamanis and Sutton~\cite{allamanis2013mining}. This corpus consists of 14,807 open-source Java projects from GitHub. To maintain consistency, we follow the dataset preprocessing protocol established in prior studies~\cite{hellendoorn2017deep, karampatsis2020big}. We extract a 1\% subset of the corpus and divide it into 12,934, 7,189, and 8,268 files for training, validation, and testing, respectively. This dataset is included in the CodeXGLUE benchmark~\cite{lu2021codexglue}, which is widely used to evaluate code completion models. We apply transformation rules to the original Java Corpus training dataset and then follow the same preprocessing steps as CodeXGLUE to enable fair comparison.

Since our objective is to examine whether license-restricted code can be exploited in fine-tuning, the transformed code is only used for fine-tuning. \revision{When evaluating the MI success rate, we use the original dataset as input, following standard MI protocols.}

Following the setup proposed by Duan \etal~\cite{duan2024membership}, we randomly select up to 1,000 samples from the GitHub Java Corpus training dataset that are applicable to a specific transformation rule as the member set. If fewer than 1,000 applicable samples exist, we include all available samples. For example, only 307 training samples satisfy Rule~4 (\textit{Do2While}), we use them as the set of members. \revision{Since our study simulates a black-box auditing scenario from the defender's perspective, we do not assume access to the model's transformed training data or knowledge of which \sect was applied. We follow a standard process to construct the non-member set by randomly sampling an equal number of examples from the GitHub Java Corpus test dataset~\cite{yang2024gotcha, duan2024membership, wan2024does}.} For each sample in both sets, we retain only code longer than 100 words and truncate each to a maximum of 200 words from the beginning, following Duan \etal~\cite{duan2024membership}.

\revision{Since different transformation rules apply to different subsets of code (e.g., only some functions contain loops or variable names that can be renamed), we do not enforce rule applicability artificially. Instead, we apply each rule exhaustively across the dataset. Importantly, for each individual transformation rule, the same member and non-member sets are used across all MI comparisons. This ensures that comparisons of MI performance reflect only the effect of the transformation rule, not differences in input samples.}

\subsection{Evaluation Metrics}

\noindent\textbf{Model performance}:
\textbf{Accuracy} is a standard metric for evaluating the performance of the model in code completion~\cite{lu2021codexglue, yang2024gotcha, wan2024does}. We follow the evaluation strategy as CodeXGLUE. For a given sample, the accuracy is calculated as the ratio of correctly predicted tokens to the total number of tokens. Dataset-level accuracy is computed by averaging accuracy scores across all test samples. This metric reflects the performance of the victim model.


\noindent\textbf{Membership inference performance}:
\textbf{AUC ROC (Area under the ROC curve)} is a widely used metric for evaluating MI performance \cite{duan2024membership, wan2024does}. MI is framed as a binary classification problem, where ground-truth labels indicate whether a sample belongs to the training set (member) or the test set (non-member). The ROC curve \revision{(receiver operating characteristic curve)} illustrates the trade-off between the true positive rate and false positive rate at various classification thresholds. AUC ROC quantifies the model’s ability to distinguish between members and non-members, with a higher AUC ROC indicating better MI performance. An AUC of 1 indicates perfect inference, while an AUC of 0.5 represents random guessing with no discriminative power.



\subsection{Average Treatment Effect Computation}
\label{sec:ate_computation}

Given that we aim to understand the real effect of performing a \revision{\sect} rule on the success of MI, we calculate \ates by formulating a causal intervention encompassing a binary treatment. \figref{fig:structural_causal_model} illustrates the \scm we devised to answer \ref{rq:causality}. More precisely, our control treatment $T_0$ (\ie null hypothesis) consists of fine-tuning and computing the MI outcomes for a given architecture
using the original member dataset (refer to \secref{sec:datasets}). Meanwhile, our treatment group $T_1$ (\ie intervention) consists of fine-tuning and computing the MI outcomes for the same architecture
using the member dataset modified by the transformation.

The potential results for MI under $T_0$ and $T_1$ come from different model fine-tuning settings
. Given that the distributions of the values for each potential outcome in the $T_0$ and $T_1$ groups originate from different fine-tuning configurations, these values are not directly comparable. To address this issue, we compute relative scores for each potential outcome in both treatments.  

To compute relative scores, we normalize each actual value by scaling it relative to the minimum and maximum values observed across all outcomes within the same treatment group. Specifically, for a given actual value $\alpha$, we compute its relative score as:  

\begin{equation}
\gamma = \frac{\alpha - \min(\mathbf{\alpha})}{\max(\mathbf{\alpha}) - \min(\mathbf{\alpha}) + \epsilon}
\label{eq:relative_score}
\end{equation}

In \equref{eq:relative_score}, $\alpha$ represents the actual value, $\mathbf{\alpha}$ denotes the set of all observed values within the treatment group, and $\epsilon$ is a small constant added to prevent division by zero. This transformation ensures that the relative scores range from 0 to 1, allowing a standardized comparison between different fine-tuning settings. The relative score formula in \equref{eq:relative_score} enables direct comparison of potential outcomes by normalizing them within a common scale.

\subsection{Implementation Details}
\label{sec:implementation}

For fine-tuning CodeGPT, we use the publicly available code and hyperparameters from CodeXGLUE. Specifically, we set the learning rate to 8e-5 and save checkpoints every 500 steps, training for a total of 5 epochs. \revision{For other models, we follow the same fine-tuning process as CodeGPT. \mrtwo{We set the learning rate to 1e-5 for \codegen, \starcoderseven, \mellum and \deepseek, and 2e-5 for the rest of the models. We fine-tuned \codellama, \codegen, and \starcoderseven for 2 epochs, and the remaining models for 3 epochs.} More details can be found in our replication package.} We select the best-performing checkpoints based on the performance in the validation data. \revision{It is worth noting that the transformed samples were reserved completely during fine-tuning. We split code files into fixed-length token blocks according to the model's context window size, except for the final segment, which may be shorter than the block size. All segments are fed into the model, meaning that the entirety of the transformed code is preserved during training. The fine-tuning and evaluation procedures are identical to those used in CodeXGLUE~\cite{lu2021codexglue}. All fine-tuned models are used for code completion, with a context window of 1024 tokens, where each token prediction is conditioned on all preceding tokens.}

\revision{We use the standard MI benchmark provided by \textsc{Mimir}~\cite{duan2024membership} without any modifications, thereby simulating a black-box setting in which the defender audits model behavior without knowledge of the specific transformations applied by the adversary during fine-tuning. Since the only difference between models is whether they were fine-tuned on transformed code, any MI changes can be attributed to the \sect.}

\section{Empirical Results}
\label{sec:results}

\begin{table*}[htbp]
    \centering
    \caption{Effect of \rulerename on MI Results (Difference in Parentheses: Transformed $-$ Original)}
    \Description{The table reports model accuracy and MI performance under the RenameVariable transformation across multiple models. Differences between transformed and original model are shown to highlight the impact on MI success rate.}
    \vspace{-1em}
    \resizebox{\textwidth}{!}{
\begin{tabular}{cccccccccc}
\hline
                           &                      & CodeGPT-small-java          & codegen-350M-multi          & deepseek-coder-1.3b          & stable-code-3b              & Mellum-4b                   & starcoder2-3b               & starcoder2-7b               & CodeLlama-7b                \\ \hline
\multirow{4}{*}{rule 1}    
& Accuracy & 74.74-76.24(-1.50) & 80.05-80.81(-0.76) & 82.07-82.70(-0.63) &80.46-81.57(-1.11) &81.50-82.65(-1.15) &80.84-82.02(-1.18) &83.79-84.01(-0.22) &83.36-83.64(-0.28) \\
& \loss & \textbf{88.88-96.87(-7.99)} & \textbf{88.16-90.79(-2.63)} & \textbf{85.17-95.36(-10.19)} & 96.62-98.91(-2.29)          & \textbf{93.58-98.45(-4.87)} & \textbf{93.71-98.52(-4.81)} & \textbf{74.69-75.88(-1.19)} & 90.56-91.47(-0.91)          \\
                           & \mink & \textbf{92.68-97.97(-5.29)} & \textbf{90.39-92.52(-2.13)} & \textbf{87.60-96.20(-8.60)}  & 97.38-99.14(-1.76)          & \textbf{94.58-98.80(-4.22)} & \textbf{94.79-98.78(-3.99)} & \textbf{77.37-78.92(-1.54)} & 91.78-92.65(-0.87)          \\
                           & \zlib & \textbf{94.39-98.06(-3.67)} & \textbf{92.59-93.43(-0.84)} & \textbf{91.59-97.02(-5.43)}  & 98.11-99.16(-1.05)          & \textbf{96.82-98.97(-2.15)} & \textbf{96.56-99.04(-2.48)} & \textbf{84.71-85.50(-0.79)} & 93.72-93.94(-0.22)          \\ \hline
\end{tabular}}

    \label{tab:rq1_RULE1}
    
\end{table*}

\begin{table*}[htbp]
    \centering
    \caption{\revision{Effect of \rulemdc on MI Results (Difference in Parentheses: Transformed $-$ Original)}}
    \Description{The table reports model accuracy and MI performance under the ModifyConstant transformation across multiple models. Differences between transformed and original model are shown to highlight the impact on MI success rate.}

     \vspace{-1em}
    \resizebox{\textwidth}{!}{
\begin{tabular}{cccccccccc}
\hline
                           &                      & CodeGPT-small-java          & codegen-350M-multi          & deepseek-coder-1.3b          & stable-code-3b              & Mellum-4b                   & starcoder2-3b               & starcoder2-7b               & CodeLlama-7b                \\ \hline
\multirow{4}{*}{rule 13}   & Accuracy & 76.19-76.24(-0.05) & 80.84-80.81(+0.03) &82.59-82.70(-0.11) &81.25-81.57(-0.32) &82.44-82.65(-0.21) &81.69-82.02(-0.33) &84.00-84.01(-0.01) &83.63-83.64(-0.01) \\
& \loss & 92.16-93.22(-1.06)          & 85.70-81.44(+4.26)          & 83.61-90.85(-7.24)           & 94.11-97.46(-3.35)          & 93.71-97.72(-4.01)          & 93.31-97.16(-3.85)          & 69.54-70.10(-0.56)          & \textbf{89.43-90.66(-1.23)} \\
                           & \mink & 95.16-95.83(-0.67)          & 88.93-84.96(+3.97)          & 86.75-92.98(-6.23)           & 95.60-98.11(-2.51)          & 95.27-98.38(-3.11)          & 95.02-97.84(-2.82)          & 72.03-72.73(-0.70)          & \textbf{91.33-92.37(-1.04)} \\
                           & \zlib & 96.38-96.89(-0.51)          & 92.34-89.87(+2.47)          & 91.07-95.36(-4.29)           & 97.12-98.94(-1.82)          & 96.87-98.99(-2.12)          & 96.50-98.57(-2.07)          & 81.24-81.82(-0.58)          & \textbf{93.48-94.32(-0.84)} \\ \hline
\end{tabular}
   }
   
     \label{tab:rq1_RULE13}
\end{table*}

\begin{table*}[htbp]
    \centering
    \caption{\revision{Effect of all rules on MI Results (Difference in Parentheses: Transformed $-$ Original)}}
    \Description{The table reports model accuracy and MI performance under all transformation across multiple models. Differences between transformed and original model are shown to highlight the impact on MI success rate.}
    
     \vspace{-1em}
    \resizebox{\textwidth}{!}{
    \begin{tabular}{cccccccccc}
\hline
                           &                      & CodeGPT-small-java          & codegen-350M-multi          & deepseek-coder-1.3b          & stable-code-3b              & Mellum-4b                   & starcoder2-3b               & starcoder2-7b               & CodeLlama-7b                \\ \hline
\multirow{4}{*}{all rules} & Accuracy & 74.65-76.24(-1.59) & 79.65-80.81(-1.16) &81.54-82.70(-1.16) &80.11-81.57(-1.46) &81.34-82.65(-1.31) &79.88-82.02(-2.14) &83.54-84.01(-0.47) &83.01-83.64(-0.63) \\
& \loss & 93.46-96.86(-3.40)          & 88.72-90.93(-2.21)          & 92.63-95.42(-2.79)           & 98.50-98.98(-0.48)          & 95.42-98.51(-3.10)          & 91.62-98.54(-6.92)          & 73.66-76.10(-2.44)          & 89.14-91.54(-2.40)          \\
                           & \mink & 96.12-97.98(-1.86)          & 91.01-92.67(-1.66)          & 94.25-96.26(-2.01)           & 98.90-99.20(-0.30)          & 96.43-98.86(-2.43)          & 92.89-98.80(-5.91)          & 76.62-79.10(-2.48)          & 90.62-92.70(-2.08)          \\
                           & \zlib & 96.96-98.10(-1.14)          & 93.17-93.47(-0.30)          & 95.65-97.12(-1.47)           & 99.05-99.26(-0.21)          & 97.60-99.06(-1.46)          & 95.27-99.09(-3.82)          & 84.08-85.50(-1.42)          & 92.78-93.98(-1.20)          \\ \hline
\end{tabular}
   }
   
    \label{tab:rq2_RULEMIX}
     \vspace{1em}
\end{table*}
\begin{table}[htbp]
\centering
\caption{\revision{Transformed Data Statistics for \ref{rq:performance_2} (\textbf{bold}: \textnormal{top 5})}}
\Description{The table shows the number and percentage of training instances affected by each transformation rule when applying combined transformations.}
\label{tab:rq2_transformed_data}
\vspace{-1em}
\scalebox{0.84}{

\begin{tabular}{cclcc}
\textbf{ID} & \textbf{Transformed Instances (\%)} &  & \textbf{ID} & \textbf{Transformed Instances (\%)} \\ \hline
1  & \textbf{ 12,143 (93.88\%)} &  & 13 & 928 (7.17\%)   \\
2  & 3,006 (23.24\%)  &  & 14 & \textbf{7,326 (56.64\%)} \\
3  & 1,650 (12.76\%)  &  & 15 & 3,296 (25.48\%) \\
4  & 286 (2.21\%)    &  & 16 & 1,689 (13.06\%) \\
5  & 2,859 (22.10\%)  &  & 17 & 3,465 (26.79\%) \\
6  & 682 (5.27\%)    &  & 18 & 648 (5.01\%)   \\
7  & 1,376 (10.64\%)  &  & 19 & 2,825 (21.84\%) \\
8  & 1,459 (11.28\%)  &  & 20 & 583 (4.51\%)   \\
9  & 909 (7.03\%)    &  & 21 & \textbf{5,721 (44.23\%)} \\
10 & 269 (2.08\%)    &  & 22 & 1,147 (8.87\%)  \\
11 & 2,425 (18.75\%)  &  & 23 & \textbf{6,317 (48.84\%)} \\
12 & \textbf{ 6,990 (54.04\%)}  &  &    &                \\ \hline
\end{tabular}

} 
\end{table}

We conducted experiments to address RQs regarding the impact of transformation rules on MI. For RQ1, we evaluate the influence of each transformation rule on model performance and the success rate of MI. For RQ2, we explore the effect of combining all transformation rules to increase their impact on MI. To answer RQ3, we examine the causal relationships between transformation rules and MI success.

\subsection{\ref{rq:performance_1}: Impact of Individual Rule on MI}

To answer RQ1, we construct a dataset for each transformation rule (described in \tabref{tab:code_transformations}) by applying the rule to the original dataset. For each rule, we apply it to all the code files applicable in the original dataset. We provide a summary of the number of transformed data instances per rule in the replication package. We then fine-tune 24 models, each trained on either the original dataset or one of the 23 transformed datasets. We measure each rule's impact on task performance. We then apply three widely used MI techniques (described in Section~\ref{sec:mi}) to assess whether an original code snippet was used for fine-tuning. 

\revision{The full experimental results are provided in the replication package. First, we observe that models fine-tuned on transformed code exhibit a comparable performance (difference up to 1.5\%) to those fine-tuned on the original data.}

\mrtwo{We present the detailed evaluation in \tabref{tab:rq1_RULE1}, where bold indicates the rule with the most significant impact on MI.} We report the accuracy of models fine-tuned in transformed datasets using \rulerename, followed by the AUC-ROC values for each MI method. We also present the difference in parentheses with negative values indicating a reduction in MI success rate. We found that \rulerename led to the most significant drop in MI performance among all \sect rules for \mrtwo{6 of the 8 models} evaluated. For the remaining three models, \rulerename still ranked consistently among the top 3 most effective transformations to reduce MI success. These results underscore the strong obfuscation effect of \rulerename on MI detection.

\revision{Specifically, on \deepseek, \rulerename reduced the \loss from 95.36\% to 85.17\% (a decrease of 10.19\%), while accuracy declined by only 0.63\%. Other MI metrics, such as \mink and \zlib, also showed substantial reductions (8.60\% and 5.43\%, respectively), further supporting the strong impact of \rulerename in multiple MI methods. A similar trend is observed on \codegpt, where \loss dropped by 7.99\% while accuracy was only slightly affected (a 1.50\% decrease). \mrtwo{For larger models like \mellum, \loss dropped by 4.87\%, substantially more than any other rule.} These findings suggest that \rulerename is highly effective in reducing MI detection with minimal impact on the utility of the model.}

We further analyze the effect of \rulemdc on MI performance. As shown in \tabref{tab:rq1_RULE13}, while the transformation slightly decreases the precision in most cases (within 0.5\%), it causes noticeable changes in MI performance across different metrics and models. In particular, on \deepseek, \rulemdc leads to a substantial drop in MI effectiveness, with \loss decreasing from 90.85\% to 83.61\% (–7.24\%), \mink decreasing by 6. 23\% and \zlib by 4.29\%, while the accuracy of the model only decreases by 0.11\%. \mrtwo{Overall, except \codegen and \starcoderseven, \rulemdc ranked consistently among the top-4 most effective transformations in reducing MI success.}

\mrtwo{Larger models demonstrate stronger resistance to the effects of \sect. When comparing \codellama and \starcoderseven with models smaller than 7B, we observe that the impact of \sect on both 7B models is below 1.54\%, substantially lower than that observed in smaller models. Furthermore, a comparison between \starcoderseven and \starcoder (\ie two models sharing the same architecture but differing in scale) shows that \rulerename remains the most influential rule, while \rulemdc exhibits moderate effects.}

\revision{Notably, we observe an exceptional trend only on \codegen on four transformations, Rules 6, 13, 18, and 22, which results in an increase in MI success rates. For example, under Rule 13, the \loss metric increases by 4.26\%, \mink by 3.97\%, and \zlib by 2.47\%. To better understand this anomalous phenomenon, we observe that other models demonstrate relatively consistent behavior—across the other 5 models, similar performance improvements are all lower than 0.69\% and without systematic patterns. In contrast, CodeGen shows more noticeable increases, suggesting the presence of structural differences.}\revision{One plausible factor is that CodeGen may differ from other models in its multi-stage sequential training strategy, which can affect the degree of redundancy or overlap encountered during training. Since repeated exposure has been linked to stronger memorization tendencies~\cite{yang2024unveiling}, our results are compatible with the view that redundancy-related effects could interact with certain \sect rules, amplifying the model’s memorization
patterns of training data. Future research could further validate this hypothesis by controlling training strategies.}

\revision{Building on the above quantitative results, we hypothesize that \rulerename and \rulemdc are particularly effective at reducing MI success because they produce code that, while functionally correct, deviates significantly from natural human coding styles. To validate this, we conducted a user study inspired by prior work~\cite{letowards}. We recruited four experienced programmers and randomly sampled 20 transformed snippets per rule per participant. We asked participants the following two questions for each code transformation: (1) ``Does the code transformation reduce code readability?'', and (2) ``Does the code transformation break the coding conventions?'' Participants responded using a 4-point Likert scale~\cite{joshi2015likert}, where 1, 2, 3, and 4 correspond to Disagree, Weakly Disagree, Weakly Agree, and Agree, respectively. Based on the average scores from all the participants, Rule 1 (top1, avg = 3.519) and Rule 13 (top4, avg = 2.875) emerged as among the most disruptive \sect, indicating that these transformations tend to produce code that significantly deviates from natural human coding practices. This may help explain why these transformations lead to notable MI performance drops. However, high-scoring rules like Rule 2 (2.955) and Rule 17 (2.905) did not reduce MI, suggesting that perceived unnaturalness alone does not determine effectiveness.}

\begin{boxK}
    \textit{\ref{rq:performance_1}}: 
    \revision{Our results show that \rulerename is a consistently effective \sect, reducing MI detection with relatively small impact on accuracy. Similarly, \rulemdc yields comparable MI reductions on several models, though its effects vary. 
    }
\end{boxK}

\subsection{\ref{rq:performance_2}: Impact of Combined Rules on MI}
To address \ref{rq:performance_2}, we first design a method to determine the order in which to apply the rules. Some transformation rules can be applied together without conflict. For example, Rule 1 (\textit{RenameVariable}) and Rule 2 (\textit{For2While}) can be easily combined by simply first replacing variable names with random strings and then converting all for loops into while loops. However, certain transformation rules are mutually exclusive. For example, Rule 2 (\textit{while-to-for}) and Rule 3 (\textit{for-to-while}) conflict with each other. To maximize rule diversity in the final transformed dataset, we first count the number of data samples that can be transformed using each rule individually. Then, we apply transformation rules in ascending order of their applicable sample size. In other words, rules with fewer applicable samples are applied first. \revision{For RQ2,} we report the number of data instance being transformed in the proposed order in \tabref{tab:rq2_transformed_data}. 
As a result, we obtain a transformed dataset to which all transformation rules were applied, then we follow the same experimental procedure as in $RQ_{1}$ to evaluate performance.

\tabref{tab:rq2_RULEMIX} presents the performance of models fine-tuned on the dataset that is applied all rules. The results show that, despite the substantial alterations introduced by applying all transformation rules, the models still maintain acceptable accuracy. \revision{Specifically, we observe that applying all transformation rules results in accuracy drops across all \mrtwo{eight} models, ranging from 0.63\% to 2.14\%—larger than the degradation caused by any individual rule. This suggests a potential trade-off: applying more transformations may amplify perturbations to code structure, thereby affecting model utility.}

The reduction in MI performance from applying all rules is not consistently greater than that of the most effective individual rule. \mrtwo{In 5 out of the 8 models, combining all transformations yields smaller drop in MI detection compared to using \rulerename alone.} For instance, on \deepseek, \loss drops by only 2.79\% when applying all rules, compared to 10.19\% with \rulerename alone. A similar pattern is observed on \codegpt (\loss: –3.40\% vs –7.99\%). This trend holds across other metrics such as \mink and \zlib. \mrtwo{Conversely, for \starcoder, \starcoderseven and \codellama, combining all rules is more effective than all the single rule-based transformation, suggesting model-specific and model-size sensitivity.} This raises an important question: under what conditions do \sect meaningfully affect MI outcomes, and are certain \sect causally responsible for reductions in MI success? To investigate this, we turn to causal analysis in RQ 3, aiming to disentangle the individual and combined effects while controlling for confounding factors.

\begin{boxK}
    \textit{\ref{rq:performance_2}}: 
    \revision{
    Fine-tuning on the dataset transformed by all 23 \sect rules results in a larger accuracy drop compared to individual transformations. However, the reduction in MI success does not surpass that of the most effective single rule (\rulerename), suggesting that \textbf{MI mitigation effects do not exhibit a cumulative pattern}.}
\end{boxK}



\subsection{\ref{rq:causality}: Causal Effects of Rules on MI}


To address \ref{rq:causality}, we compute the \ates of each transformation rule on the effectiveness of MI for each potential outcome, as depicted in the proposed \scm (refer to \secref{sec:ate_computation}). \revision{Results are summarized in \tabref{tab:causal_results_2} and \tabref{tab:causal_results_1}, covering \mrtwo{eight} models spanning diverse scales and architectures.}


\revision{Across all models, \textbf{Rule 1 (RenameVariable)} consistently demonstrates a strong \textit{negative} causal effect on MI effectiveness, particularly when measured by \mink. Specifically, Rule 1 reduces \mink by $-0.78$ in $M_A$ (\codegen), $-0.05$ in $M_B$ (\codegpt), $-0.23$ in $M_C$ (\codellama), $-0.83$ in $M_D$ (\deepseek), $-0.76$ in $M_E$ (\starcoder), \mrtwo{$-0.16$ in $M_G$ (\starcoderseven), $-0.75$ in $M_F$ (\stablecode), and $-0.92$ in $M_H$ (\mellum). Corresponding effects on \loss are also consistently negative across these models.}}

\begin{table*}[]

\centering
\caption{\revision{Causal Analysis (\ate) results for  $M_{D-H}$}}
\Description{The table reports average treatment effects from the causal analysis for several models, measuring the impact of individual transformation rules on different membership inference metrics.}
\label{tab:causal_results_2}
\vspace{-1em}
\scalebox{0.54}{

\setlength{\tabcolsep}{4pt} 

\begin{tabular}{lllllllllllllllllllllllllllllllllllllll}
\multicolumn{7}{c}{$M_D$ - \deepseek \cite{guo2024deepseek}} &
  \multicolumn{1}{c}{} &
  \multicolumn{7}{c}{$M_E$ - \starcoder \cite{li2023starcoder}} &
  \multicolumn{1}{c}{} &
  \multicolumn{7}{c}{$M_G$ - \starcoderseven \cite{li2023starcoder}} &
  \multicolumn{1}{c}{} &
  \multicolumn{7}{c}{$M_F$ - \stablecode \cite{stable-code-3b}} &
  \multicolumn{1}{c}{} &
  \multicolumn{7}{c}{$M_H$ - \mellum \cite{Mellum-4b-base}} \\
\multicolumn{1}{c}{} &
  \multicolumn{2}{c}{\textit{\textbf{\loss}}} &
  \multicolumn{2}{c}{\textit{\textbf{\zlib}}} &
  \multicolumn{2}{c}{\textit{\textbf{\mink}}} &
   &
  \multicolumn{1}{c}{} &
  \multicolumn{2}{c}{\textit{\textbf{\loss}}} &
  \multicolumn{2}{c}{\textit{\textbf{\zlib}}} &
  \multicolumn{2}{c}{\textit{\textbf{\mink}}} &
   &
   &
  \multicolumn{2}{c}{\textit{\textbf{\loss}}} &
  \multicolumn{2}{c}{\textit{\textbf{\zlib}}} &
  \multicolumn{2}{c}{\textit{\textbf{\mink}}} &
  \multicolumn{1}{c}{\textit{\textbf{}}} &
  \multicolumn{1}{c}{\textit{\textbf{}}} &
  \multicolumn{2}{c}{\textit{\textbf{\loss}}} &
  \multicolumn{2}{c}{\textit{\textbf{\zlib}}} &
  \multicolumn{2}{c}{\textit{\textbf{\mink}}} &
   &
   &
  \multicolumn{2}{c}{\textit{\textbf{\loss}}} &
  \multicolumn{2}{c}{\textit{\textbf{\zlib}}} &
  \multicolumn{2}{c}{\textit{\textbf{\mink}}} \\
\multicolumn{1}{c}{rule} &
  \multicolumn{1}{c}{\textit{p}} &
  \multicolumn{1}{c}{\textit{\ate}} &
  \multicolumn{1}{c}{\textit{p}} &
  \multicolumn{1}{c}{\textit{\ate}} &
  \multicolumn{1}{c}{\textit{p}} &
  \multicolumn{1}{c}{\textit{\ate}} &
   &
  \multicolumn{1}{c}{rule} &
  \multicolumn{1}{c}{\textit{p}} &
  \multicolumn{1}{c}{\textit{\ate}} &
  \multicolumn{1}{c}{\textit{p}} &
  \multicolumn{1}{c}{\textit{\ate}} &
  \multicolumn{1}{c}{\textit{p}} &
  \multicolumn{1}{c}{\textit{\ate}} &
   &
  \multicolumn{1}{c}{rule} &
  \multicolumn{1}{c}{\textit{p}} &
  \multicolumn{1}{c}{\textit{\ate}} &
  \multicolumn{1}{c}{\textit{p}} &
  \multicolumn{1}{c}{\textit{\ate}} &
  \multicolumn{1}{c}{\textit{p}} &
  \multicolumn{1}{c}{\textit{\ate}} &
   &
  \multicolumn{1}{c}{rule} &
  \multicolumn{1}{c}{\textit{p}} &
  \multicolumn{1}{c}{\textit{\ate}} &
  \multicolumn{1}{c}{\textit{p}} &
  \multicolumn{1}{c}{\textit{\ate}} &
  \multicolumn{1}{c}{\textit{p}} &
  \multicolumn{1}{c}{\textit{\ate}} &
   &
  \multicolumn{1}{c}{rule} &
  \multicolumn{1}{c}{\textit{p}} &
  \multicolumn{1}{c}{\textit{\ate}} &
  \multicolumn{1}{c}{\textit{p}} &
  \multicolumn{1}{c}{\textit{\ate}} &
  \multicolumn{1}{c}{\textit{p}} &
  \multicolumn{1}{c}{\textit{\ate}} \\ \cline{1-7} \cline{9-15} \cline{17-23} \cline{25-31} \cline{33-39} 
1 &
  {\ul \textbf{-0.52}} &
  \cellcolor[HTML]{6D739C}{\color[HTML]{FFFFFF} -0.43} &
  {\ul \textbf{-0.48}} &
  0 &
  {\ul \textbf{-0.55}} &
  \cellcolor[HTML]{6D739C}{\color[HTML]{FFFFFF} -0.83} &
   &
  1 &
  {\ul \textbf{-0.5}} &
  \cellcolor[HTML]{6D739C}{\color[HTML]{FFFFFF} -0.34} &
  {\ul \textbf{-0.50}} &
  0 &
  {\ul \textit{\textbf{-0.54}}} &
  \cellcolor[HTML]{6D739C}{\color[HTML]{FFFFFF} -0.76} &
   &
  1 &
  -0.07 &
  -0.04 &
  -0.06 &
  0 &
  -0.08 &
  \cellcolor[HTML]{6D739C}{\color[HTML]{FFFFFF} -0.16} &
   &
  1 &
  {\ul \textbf{-0.46}} &
  \cellcolor[HTML]{6D739C}{\color[HTML]{FFFFFF} -0.16} &
  {\ul \textbf{-0.41}} &
  0 &
  {\ul \textbf{-0.49}} &
  \cellcolor[HTML]{6D739C}{\color[HTML]{FFFFFF} -0.75} &
   &
  1 &
  {\ul \textbf{-0.54}} &
  \cellcolor[HTML]{6D739C}{\color[HTML]{FFFFFF} -0.42} &
  {\ul \textit{\textbf{-0.52}}} &
  0 &
  {\ul \textbf{-0.56}} &
  \cellcolor[HTML]{6D739C}{\color[HTML]{FFFFFF} -0.92} \\
2 &
  {\ul \textbf{-0.44}} &
  \cellcolor[HTML]{6D739C}{\color[HTML]{FFFFFF} -0.18} &
  {\ul \textbf{-0.46}} &
  0 &
  {\ul \textbf{-0.46}} &
  \cellcolor[HTML]{6D739C}{\color[HTML]{FFFFFF} -0.79} &
   &
  7 &
  {\ul \textbf{-0.35}} &
  \cellcolor[HTML]{6D739C}{\color[HTML]{FFFFFF} -0.17} &
  \textbf{-0.09} &
  0 &
  \textit{\textbf{-0.36}} &
  \cellcolor[HTML]{6D739C}{\color[HTML]{FFFFFF} -0.73} &
   &
  23 &
  -0.02 &
  -0.01 &
  -0.02 &
  0 &
  -0.02 &
  -0.05 &
   &
  2 &
  {\ul \textbf{-0.10}} &
  -0.03 &
  {\ul \textbf{-0.10}} &
  0 &
  {\ul \textbf{-0.11}} &
  \cellcolor[HTML]{6D739C}{\color[HTML]{FFFFFF} -0.16} &
   &
  4 &
  {\ul \textbf{-0.57}} &
  \cellcolor[HTML]{6D739C}{\color[HTML]{FFFFFF} -0.26} &
  {\ul \textit{\textbf{-0.57}}} &
  0 &
  {\ul \textbf{-0.58}} &
  \cellcolor[HTML]{6D739C}{\color[HTML]{FFFFFF} -0.92} \\
6 &
  {\ul \textbf{-0.29}} &
  \cellcolor[HTML]{6D739C}{\color[HTML]{FFFFFF} -0.15} &
  {\ul \textbf{-0.31}} &
  0 &
  {\ul \textbf{-0.31}} &
  \cellcolor[HTML]{6D739C}{\color[HTML]{FFFFFF} -0.64} &
   &
  12 &
  {\ul \textbf{-0.25}} &
  \cellcolor[HTML]{6D739C}{\color[HTML]{FFFFFF} -0.12} &
  {\ul \textbf{-0.26}} &
  0 &
  {\ul \textit{\textbf{-0.25}}} &
  \cellcolor[HTML]{6D739C}{\color[HTML]{FFFFFF} -0.51} &
   &
  14 &
  -0.03 &
  -0.01 &
  -0.03 &
  0 &
  -0.03 &
  -0.05 &
   &
  3 &
  {\ul \textbf{-0.36}} &
  \cellcolor[HTML]{6D739C}{\color[HTML]{FFFFFF} -0.17} &
  {\ul \textbf{-0.37}} &
  0 &
  {\ul \textbf{-0.38}} &
  \cellcolor[HTML]{6D739C}{\color[HTML]{FFFFFF} -0.76} &
   &
  18 &
  {\ul \textbf{-0.36}} &
  \cellcolor[HTML]{6D739C}{\color[HTML]{FFFFFF} -0.16} &
  {\ul \textit{\textbf{-0.36}}} &
  0 &
  {\ul \textbf{-0.37}} &
  \cellcolor[HTML]{6D739C}{\color[HTML]{FFFFFF} -0.69} \\
8 &
  {\ul \textbf{-0.25}} &
  \cellcolor[HTML]{6D739C}{\color[HTML]{FFFFFF} -0.17} &
  {\ul \textbf{-0.26}} &
  0 &
  {\ul \textbf{-0.25}} &
  \cellcolor[HTML]{6D739C}{\color[HTML]{FFFFFF} -0.47} &
   &
  13 &
  {\ul \textbf{-0.29}} &
  \cellcolor[HTML]{6D739C}{\color[HTML]{FFFFFF} -0.14} &
  {\ul \textbf{-0.29}} &
  0 &
  \textit{\textbf{-0.29}} &
  \cellcolor[HTML]{6D739C}{\color[HTML]{FFFFFF} -0.57} &
   &
  16 &
  -0.01 &
  -0.01 &
  0 &
  0 &
  -0.01 &
  -0.03 &
   &
  6 &
  {\ul \textbf{-0.34}} &
  \cellcolor[HTML]{6D739C}{\color[HTML]{FFFFFF} -0.16} &
  {\ul \textbf{-0.36}} &
  0 &
  {\ul \textbf{-0.34}} &
  \cellcolor[HTML]{6D739C}{\color[HTML]{FFFFFF} -0.66} &
   &
  16 &
  {\ul \textbf{-0.42}} &
  \cellcolor[HTML]{6D739C}{\color[HTML]{FFFFFF} -0.15} &
  {\ul \textit{\textbf{-0.42}}} &
  0 &
  {\ul \textbf{-0.42}} &
  \cellcolor[HTML]{6D739C}{\color[HTML]{FFFFFF} -0.64} \\
10 &
  {\ul \textbf{-0.21}} &
  \cellcolor[HTML]{6D739C}{\color[HTML]{FFFFFF} -0.11} &
  {\ul \textbf{-0.23}} &
  0 &
  {\ul \textbf{-0.23}} &
  \cellcolor[HTML]{6D739C}{\color[HTML]{FFFFFF} -0.47} &
   &
  14 &
  -0.09 &
  \multicolumn{1}{r}{-0.04} &
  \multicolumn{1}{r}{-0.09} &
  0 &
  \textit{\textbf{-0.11}} &
  \cellcolor[HTML]{6D739C}{\color[HTML]{FFFFFF} -0.20} &
   &
  4 &
  -0.01 &
  -0.01 &
  0 &
  0 &
  0 &
  -0.02 &
   &
  9 &
  {\ul \textbf{-0.33}} &
  \cellcolor[HTML]{6D739C}{\color[HTML]{FFFFFF} -0.16} &
  {\ul \textbf{-0.34}} &
  0 &
  {\ul \textbf{-0.35}} &
  \cellcolor[HTML]{6D739C}{\color[HTML]{FFFFFF} -0.68} &
   &
  6 &
  {\ul \textbf{-0.30}} &
  \cellcolor[HTML]{6D739C}{\color[HTML]{FFFFFF} -0.14} &
  {\ul \textit{\textbf{-0.32}}} &
  0 &
  {\ul \textbf{-0.31}} &
  \cellcolor[HTML]{6D739C}{\color[HTML]{FFFFFF} -0.61} \\
11 &
  {\ul \textbf{-0.30}} &
  \cellcolor[HTML]{6D739C}{\color[HTML]{FFFFFF} -0.12} &
  {\ul \textbf{-0.32}} &
  0 &
  {\ul \textbf{-0.32}} &
  \cellcolor[HTML]{6D739C}{\color[HTML]{FFFFFF} -0.52} &
   &
  15 &
  {\ul \textbf{-0.29}} &
  \multicolumn{1}{r}{\cellcolor[HTML]{6D739C}{\color[HTML]{FFFFFF} -0.12}} &
  {\ul \textbf{-0.26}} &
  0 &
  \textit{\textbf{-0.29}} &
  \cellcolor[HTML]{6D739C}{\color[HTML]{FFFFFF} -0.51} &
   &
  8 &
  0 &
  0.01 &
  0.01 &
  0 &
  0 &
  0.03 &
   &
  10 &
  -0.07 &
  -0.03 &
  -0.08 &
  0 &
  -0.09 &
  \cellcolor[HTML]{6D739C}{\color[HTML]{FFFFFF} -0.16} &
   &
  2 &
  {\ul \textbf{-0.39}} &
  \cellcolor[HTML]{6D739C}{\color[HTML]{FFFFFF} -0.14} &
  {\ul \textit{\textbf{-0.41}}} &
  0 &
  {\ul \textbf{-0.39}} &
  \cellcolor[HTML]{6D739C}{\color[HTML]{FFFFFF} -0.60} \\
13 &
  {\ul \textbf{-0.34}} &
  \cellcolor[HTML]{6D739C}{\color[HTML]{FFFFFF} -0.16} &
  {\ul \textbf{-0.35}} &
  0 &
  {\ul \textbf{-0.38}} &
  \cellcolor[HTML]{6D739C}{\color[HTML]{FFFFFF} -0.71} &
   &
  17 &
  {\ul \textbf{-0.27}} &
  \cellcolor[HTML]{6D739C}{\color[HTML]{FFFFFF} -0.12} &
  {\ul \textbf{-0.26}} &
  0 &
  \textit{\textbf{-0.27}} &
  \cellcolor[HTML]{6D739C}{\color[HTML]{FFFFFF} -0.54} &
   &
  3 &
  -0.01 &
  0.01 &
  0 &
  0 &
  0 &
  0.03 &
   &
  12 &
  -0.07 &
  -0.03 &
  -0.07 &
  0 &
  -0.09 &
  \cellcolor[HTML]{6D739C}{\color[HTML]{FFFFFF} -0.14} &
   &
  11 &
  {\ul \textbf{-0.37}} &
  \cellcolor[HTML]{6D739C}{\color[HTML]{FFFFFF} -0.13} &
  {\ul \textit{\textbf{-0.38}}} &
  0 &
  {\ul \textbf{-0.38}} &
  \cellcolor[HTML]{6D739C}{\color[HTML]{FFFFFF} -0.58} \\
14 &
  {\ul \textbf{-0.25}} &
  \cellcolor[HTML]{6D739C}{\color[HTML]{FFFFFF} -0.12} &
  {\ul \textbf{-0.25}} &
  0 &
  {\ul \textbf{-0.27}} &
  \cellcolor[HTML]{6D739C}{\color[HTML]{FFFFFF} -0.52} &
   &
  20 &
  {\ul \textbf{-0.33}} &
  \cellcolor[HTML]{6D739C}{\color[HTML]{FFFFFF} -0.15} &
  {\ul \textbf{-0.36}} &
  0 &
  \textbf{-0.34} &
  \cellcolor[HTML]{6D739C}{\color[HTML]{FFFFFF} -0.64} &
   &
  7 &
  -0.01 &
  0.02 &
  -0.01 &
  0 &
  -0.01 &
  0.06 &
   &
  13 &
  {\ul \textbf{-0.31}} &
  \cellcolor[HTML]{6D739C}{\color[HTML]{FFFFFF} -0.15} &
  {\ul \textbf{-0.31}} &
  0 &
  {\ul \textbf{-0.33}} &
  \cellcolor[HTML]{6D739C}{\color[HTML]{FFFFFF} -0.69} &
   &
  9 &
  {\ul \textbf{-0.29}} &
  \cellcolor[HTML]{6D739C}{\color[HTML]{FFFFFF} -0.13} &
  {\ul \textit{\textbf{-0.29}}} &
  0 &
  {\ul \textbf{-0.29}} &
  \cellcolor[HTML]{6D739C}{\color[HTML]{FFFFFF} -0.51} \\
16 &
  {\ul \textbf{-0.13}} &
  -0.04 &
  {\ul \textbf{-0.13}} &
  0 &
  {\ul \textbf{-0.14}} &
  \cellcolor[HTML]{6D739C}{\color[HTML]{FFFFFF} -0.18} &
   &
  21 &
  {\ul \textbf{-0.26}} &
  \cellcolor[HTML]{6D739C}{\color[HTML]{FFFFFF} -0.12} &
  {\ul \textbf{-0.26}} &
  0 &
  \textit{\textbf{-0.26}} &
  \cellcolor[HTML]{6D739C}{\color[HTML]{FFFFFF} -0.53} &
   &
  ALL &
  {\ul \textbf{-0.12}} &
  -0.05 &
  {\ul \textbf{-0.11}} &
  0 &
  {\ul \textbf{-0.13}} &
  \cellcolor[HTML]{6D739C}{\color[HTML]{FFFFFF} -0.20} &
   &
  16 &
  {\ul \textbf{-0.41}} &
  \cellcolor[HTML]{6D739C}{\color[HTML]{FFFFFF} -0.15} &
  {\ul \textbf{-0.42}} &
  0 &
  {\ul \textbf{-0.42}} &
  \cellcolor[HTML]{6D739C}{\color[HTML]{FFFFFF} -0.66} &
   &
  13 &
  {\ul \textbf{-0.29}} &
  \cellcolor[HTML]{6D739C}{\color[HTML]{FFFFFF} -0.12} &
  {\ul \textit{\textbf{-0.28}}} &
  0 &
  {\ul \textbf{-0.27}} &
  \cellcolor[HTML]{6D739C}{\color[HTML]{FFFFFF} -0.49} \\ \cline{17-23}
17 &
  {\ul \textbf{-0.30}} &
  \cellcolor[HTML]{6D739C}{\color[HTML]{FFFFFF} -0.13} &
  {\ul \textbf{-0.30}} &
  0 &
  {\ul \textbf{-0.31}} &
  \cellcolor[HTML]{6D739C}{\color[HTML]{FFFFFF} -0.58} &
   &
  22 &
  {\ul \textbf{-0.26}} &
  \cellcolor[HTML]{6D739C}{\color[HTML]{FFFFFF} -0.13} &
  \multicolumn{1}{r}{{\ul \textbf{-0.26}}} &
  0 &
  {\ul \textbf{-0.25}} &
  \cellcolor[HTML]{6D739C}{\color[HTML]{FFFFFF} -0.53} &
   &
   &
   &
   &
   &
   &
   &
   &
   &
  17 &
  {\ul \textbf{-0.33}} &
  \cellcolor[HTML]{6D739C}{\color[HTML]{FFFFFF} -0.14} &
  {\ul \textbf{-0.33}} &
  0 &
  {\ul \textbf{-0.34}} &
  \cellcolor[HTML]{6D739C}{\color[HTML]{FFFFFF} -0.62} &
   &
  5 &
  {\ul \textbf{-0.35}} &
  \cellcolor[HTML]{6D739C}{\color[HTML]{FFFFFF} -0.12} &
  {\ul \textit{\textbf{-0.35}}} &
  0 &
  {\ul \textbf{-0.35}} &
  \cellcolor[HTML]{6D739C}{\color[HTML]{FFFFFF} -0.54} \\
21 &
  {\ul \textbf{-0.27}} &
  \cellcolor[HTML]{6D739C}{\color[HTML]{FFFFFF} -0.12} &
  {\ul \textbf{-0.26}} &
  0 &
  {\ul \textbf{-0.27}} &
  \cellcolor[HTML]{6D739C}{\color[HTML]{FFFFFF} -0.51} &
   &
  23 &
  {\ul \textbf{-0.26}} &
  \cellcolor[HTML]{6D739C}{\color[HTML]{FFFFFF} -0.13} &
  {\ul \textbf{-0.26}} &
  0 &
  \textbf{-0.27} &
  \cellcolor[HTML]{6D739C}{\color[HTML]{FFFFFF} -0.55} &
   &
   &
   &
   &
   &
   &
   &
   &
   &
  18 &
  {\ul \textbf{-0.39}} &
  \cellcolor[HTML]{6D739C}{\color[HTML]{FFFFFF} -0.17} &
  {\ul \textbf{-0.39}} &
  0 &
  {\ul \textbf{-0.41}} &
  \cellcolor[HTML]{6D739C}{\color[HTML]{FFFFFF} -0.73} &
   &
  20 &
  {\ul \textbf{-0.32}} &
  \cellcolor[HTML]{6D739C}{\color[HTML]{FFFFFF} -0.12} &
  {\ul \textit{\textbf{-0.33}}} &
  0 &
  {\ul \textbf{-0.30}} &
  \cellcolor[HTML]{6D739C}{\color[HTML]{FFFFFF} -0.48} \\
\textit{ALL} &
  {\ul \textbf{-0.38}} &
  \cellcolor[HTML]{6D739C}{\color[HTML]{FFFFFF} -0.15} &
  {\ul \textbf{-0.35}} &
  0 &
  {\ul \textbf{-0.41}} &
  \cellcolor[HTML]{6D739C}{\color[HTML]{FFFFFF} -0.68} &
   &
  ALL &
  {\ul \textbf{-0.61}} &
  \cellcolor[HTML]{6D739C}{\color[HTML]{FFFFFF} -0.35} &
  {\ul \textbf{-0.60}} &
  0 &
  {\ul \textbf{-0.63}} &
  \cellcolor[HTML]{6D739C}{\color[HTML]{FFFFFF} -0.89} &
   &
   &
   &
   &
   &
   &
   &
   &
   &
  19 &
  {\ul \textbf{-0.12}} &
  -0.03 &
  {\ul \textbf{-0.12}} &
  0 &
  {\ul \textbf{-0.13}} &
  \cellcolor[HTML]{6D739C}{\color[HTML]{FFFFFF} -0.16} &
   &
  3 &
  {\ul \textbf{-0.28}} &
  \cellcolor[HTML]{6D739C}{\color[HTML]{FFFFFF} -0.12} &
  {\ul \textit{\textbf{-0.28}}} &
  0 &
  {\ul \textbf{-0.27}} &
  \cellcolor[HTML]{6D739C}{\color[HTML]{FFFFFF} -0.49} \\ \cline{1-7} \cline{9-15}
 &
   &
   &
   &
   &
   &
   &
   &
   &
  \multicolumn{1}{r}{\textbf{}} &
   &
  \textbf{} &
   &
   &
   &
   &
   &
   &
   &
   &
   &
   &
   &
   &
  20 &
  {\ul \textbf{-0.32}} &
  \cellcolor[HTML]{6D739C}{\color[HTML]{FFFFFF} -0.13} &
  {\ul \textbf{-0.34}} &
  0 &
  {\ul \textbf{-0.37}} &
  \cellcolor[HTML]{6D739C}{\color[HTML]{FFFFFF} -0.58} &
   &
  22 &
  {\ul \textbf{-0.27}} &
  \cellcolor[HTML]{6D739C}{\color[HTML]{FFFFFF} -0.11} &
  {\ul \textit{\textbf{-0.27}}} &
  0 &
  {\ul \textbf{-0.25}} &
  \cellcolor[HTML]{6D739C}{\color[HTML]{FFFFFF} -0.43} \\
 &
   &
   &
   &
   &
   &
   &
   &
   &
  \textbf{} &
   &
  \textbf{} &
   &
  \textbf{} &
   &
   &
   &
   &
   &
   &
   &
   &
   &
   &
  \textit{ALL} &
  {\ul \textbf{-0.42}} &
  \cellcolor[HTML]{6D739C}{\color[HTML]{FFFFFF} -0.17} &
  \multicolumn{1}{r}{{\ul \textbf{-0.39}}} &
  0 &
  {\ul \textbf{-0.46}} &
  \cellcolor[HTML]{6D739C}{\color[HTML]{FFFFFF} -0.80} &
   &
  ALL &
  {\ul \textbf{-0.52}} &
  \cellcolor[HTML]{6D739C}{\color[HTML]{FFFFFF} -0.22} &
  {\ul \textit{\textbf{-0.50}}} &
  0 &
  {\ul \textbf{-0.55}} &
  \cellcolor[HTML]{6D739C}{\color[HTML]{FFFFFF} -0.90} \\ \cline{25-31} \cline{33-39} 
\end{tabular}

} 
\caption*{\small{bold: \textnormal{$-$ correlation}, \underline{bold underlined}: \textnormal{$+$ correlation}, {\color[HTML]{6D739C} background blue}: \textnormal{$-$ causal effect}, {\color[HTML]{6CB0B1} background \revision{green}}: \textnormal{$+$ causal effect}}}
\Description{The table reports average treatment effects from the causal analysis for several models, measuring the impact of individual transformation rules on different membership inference metrics.}

\vspace{-1.7em}
\end{table*}
\begin{table}[]

\centering
\caption{\revision{Causal Analysis (\ate) results for  $M_{A-D}$}}
\label{tab:causal_results_1}
\vspace{-1em}
\scalebox{0.65}{

\setlength{\tabcolsep}{4pt} 

\begin{tabular}{lrrrrrrllrrrrrr}
\multicolumn{7}{c}{$M_A$ - \codegen \cite{nijkamp2023codegen}} &
   &
  \multicolumn{7}{c}{$M_B$ - \codegpt \cite{lu2021codexglue}} \\
\multicolumn{1}{c}{} &
  \multicolumn{2}{c}{\textit{\textbf{\loss}}} &
  \multicolumn{2}{c}{\textit{\textbf{\zlib}}} &
  \multicolumn{2}{c}{\textit{\textbf{\mink}}} &
  \multicolumn{1}{c}{} &
  \multicolumn{1}{c}{} &
  \multicolumn{2}{c}{\textit{\textbf{\loss}}} &
  \multicolumn{2}{c}{\textit{\textbf{\zlib}}} &
  \multicolumn{2}{c}{\textit{\textbf{\mink}}} \\
\multicolumn{1}{c}{rule} &
  \multicolumn{1}{c}{\textit{p}} &
  \multicolumn{1}{c}{\textit{\ate}} &
  \multicolumn{1}{c}{\textit{p}} &
  \multicolumn{1}{c}{\textit{\ate}} &
  \multicolumn{1}{c}{\textit{p}} &
  \multicolumn{1}{c}{\textit{\ate}} &
  \multicolumn{1}{c}{} &
  \multicolumn{1}{c}{rule} &
  \multicolumn{1}{c}{\textit{p}} &
  \multicolumn{1}{c}{\textit{\ate}} &
  \multicolumn{1}{c}{\textit{p}} &
  \multicolumn{1}{c}{\textit{\ate}} &
  \multicolumn{1}{c}{\textit{p}} &
  \multicolumn{1}{c}{\textit{\ate}} \\ \cline{1-7} \cline{9-15} 
\textit{1} &
  \textbf{-0.23} &
  \cellcolor[HTML]{6D739C}{\color[HTML]{FFFFFF} -0.17} &
  -0.2 &
  0 &
  \textbf{-0.27} &
  \cellcolor[HTML]{6D739C}{\color[HTML]{FFFFFF} -0.78} &
  \multicolumn{1}{c}{} &
  \textit{1} &
  -0.02 &
  -0.08 &
  -0.02 &
  0 &
  -0.02 &
  -0.05 \\
\textit{2} &
  -0.04 &
  -0.03 &
  -0.05 &
  0 &
  -0.04 &
  \cellcolor[HTML]{6D739C}{\color[HTML]{FFFFFF} -0.12} &
  \multicolumn{1}{c}{} &
  \textit{13} &
  -0.05 &
  -0.03 &
  -0.05 &
  0 &
  -0.04 &
  \cellcolor[HTML]{6D739C}{\color[HTML]{FFFFFF} -0.11} \\
\textit{8} &
  -0.10 &
  -0.03 &
  \textbf{-0.11} &
  0 &
  \textbf{-0.12} &
  \cellcolor[HTML]{6D739C}{\color[HTML]{FFFFFF} -0.16} &
  \multicolumn{1}{c}{} &
  \textit{18} &
  -0.08 &
  -0.05 &
  -0.08 &
  0 &
  -0.01 &
  \cellcolor[HTML]{6D739C}{\color[HTML]{FFFFFF} -0.19} \\
\textit{10} &
  -0.01 &
  0.04 &
  -0.01 &
  0 &
  0 &
  \cellcolor[HTML]{6CB0B1}{\color[HTML]{FFFFFF} 0.14} &
  \multicolumn{1}{c}{} &
  \textit{20} &
  {\ul \textbf{0.04}} &
  0.04 &
  {\color[HTML]{333333} \textbf{0.04}} &
  0 &
  {\ul \textbf{0.04}} &
  \cellcolor[HTML]{6CB0B1}{\color[HTML]{FFFFFF} 0.13} \\
\textit{13} &
  0.09 &
  0.05 &
  0.09 &
  0 &
  0.09 &
  \cellcolor[HTML]{6CB0B1}{\color[HTML]{FFFFFF} 0.21} &
   &
  \textit{ALL} &
  {\ul \textbf{-0.30}} &
  \cellcolor[HTML]{6D739C}{\color[HTML]{FFFFFF} -0.31} &
  {\color[HTML]{333333} \textbf{-0.29}} &
  0 &
  {\ul \textbf{-0.3}} &
  \cellcolor[HTML]{6D739C}{\color[HTML]{FFFFFF} -0.85} \\ \cline{9-15} 
\textit{14} &
  -0.05 &
  -0.02 &
  -0.05 &
  0 &
  -0.05 &
  \cellcolor[HTML]{6D739C}{\color[HTML]{FFFFFF} -0.11} &
   &
   &
  \multicolumn{1}{l}{} &
  \multicolumn{1}{l}{} &
  \multicolumn{1}{l}{} &
  \multicolumn{1}{l}{} &
  \multicolumn{1}{l}{} &
  \multicolumn{1}{l}{} \\
\textit{22} &
  {\ul \textbf{0.10}} &
  0.06 &
  {\color[HTML]{333333} 0.10} &
  0 &
  {\ul \textbf{0.10}} &
  \cellcolor[HTML]{6CB0B1}{\color[HTML]{FFFFFF} 0.22} &
   &
  \multicolumn{7}{c}{$M_C$ - \codellama \cite{roziere2023code}} \\
\textit{23} &
  \textbf{-0.13} &
  -0.06 &
  {\color[HTML]{333333} -0.13} &
  0 &
  \textbf{-0.15} &
  \cellcolor[HTML]{6D739C}{\color[HTML]{FFFFFF} -0.26} &
   &
  \multicolumn{1}{c}{{\color[HTML]{D4D4D4} }} &
  \multicolumn{2}{c}{\textit{\textbf{\loss}}} &
  \multicolumn{2}{c}{\textit{\textbf{\zlib}}} &
  \multicolumn{2}{c}{\textit{\textbf{\mink}}} \\
\textit{ALL} &
  \textbf{-0.22} &
  \cellcolor[HTML]{6D739C}{\color[HTML]{FFFFFF} -0.18} &
  {\color[HTML]{333333} -0.19} &
  0 &
  \textbf{-0.24} &
  \cellcolor[HTML]{6D739C}{\color[HTML]{FFFFFF} -0.75} &
   &
  \multicolumn{1}{c}{rule} &
  \multicolumn{1}{c}{{\ul \textit{\textbf{p}}}} &
  \multicolumn{1}{c}{\textit{\textbf{\ate}}} &
  \multicolumn{1}{c}{{\ul \textit{\textbf{p}}}} &
  \multicolumn{1}{c}{\textit{\textbf{\ate}}} &
  \multicolumn{1}{c}{{\ul \textit{\textbf{p}}}} &
  \multicolumn{1}{c}{\textit{\ate}} \\ \cline{1-7} \cline{9-15} 
 &
  \multicolumn{1}{l}{} &
  \multicolumn{1}{l}{} &
  \multicolumn{1}{l}{} &
  \multicolumn{1}{l}{} &
  \multicolumn{1}{l}{} &
  \multicolumn{1}{l}{} &
   &
  1 &
  \multicolumn{1}{l}{{\ul \textit{\textbf{-0.16}}}} &
  \multicolumn{1}{l}{\textit{-0.05}} &
  \multicolumn{1}{l}{{\ul \textbf{-0.13}}} &
  \multicolumn{1}{l}{\textit{0}} &
  \multicolumn{1}{l}{{\ul \textit{\textbf{-0.18}}}} &
  \multicolumn{1}{l}{\cellcolor[HTML]{6D739C}{\color[HTML]{FFFFFF} \textit{-0.23}}} \\
 &
  \multicolumn{1}{l}{} &
  \multicolumn{1}{l}{} &
  \multicolumn{1}{l}{} &
  \multicolumn{1}{l}{} &
  \multicolumn{1}{l}{} &
  \multicolumn{1}{l}{} &
   &
  14 &
  \multicolumn{1}{l}{{\ul \textbf{-0.06}}} &
  \multicolumn{1}{l}{-0.02} &
  \multicolumn{1}{l}{{\ul \textbf{0.06}}} &
  \multicolumn{1}{l}{0} &
  \multicolumn{1}{l}{{\ul \textit{\textbf{-0.07}}}} &
  \multicolumn{1}{l}{\cellcolor[HTML]{6D739C}{\color[HTML]{FFFFFF} -0.12}} \\
 &
  \multicolumn{1}{l}{} &
  \multicolumn{1}{l}{} &
  \multicolumn{1}{l}{} &
  \multicolumn{1}{l}{} &
  \multicolumn{1}{l}{} &
  \multicolumn{1}{l}{} &
   &
  ALL &
  \multicolumn{1}{l}{{\ul \textbf{-0.23}}} &
  \multicolumn{1}{l}{-0.07} &
  \multicolumn{1}{l}{{\ul \textbf{-0.19}}} &
  \multicolumn{1}{l}{0} &
  \multicolumn{1}{l}{{\ul \textit{\textbf{-0.24}}}} &
  \multicolumn{1}{l}{\cellcolor[HTML]{6D739C}{\color[HTML]{FFFFFF} -0.29}} \\ \cline{9-15} 
\end{tabular}

} 
\caption*{\small{bold: \textnormal{$-$ correlation}, \underline{bold underlined}: \textnormal{$+$ correlation}, {\color[HTML]{6D739C} background blue}: \textnormal{$-$ causal effect}, {\color[HTML]{6CB0B1} background \revision{green}}: \textnormal{$+$ causal effect}}}
\vspace{-2em}
\end{table}

\mrtwo{The comparison between $M_E$ (\starcoder) and $M_G$ (\starcoderseven) reveals that the \ate values remain close to zero across all transformation rules, indicating negligible causal impact. \ates near zero suggest that increasing model size does not meaningfully enhance or reduce MI effectiveness under the applied transformations. Although $M_G$ exhibits marginally higher \ates than $M_E$, the differences are minor and statistically insignificant.}

\revision{Furthermore, \textbf{Rule 13 (ReorderStatements)} exhibits significant negative effects on \mink, with \ate values of $-0.57$ in $M_E$ (\starcoder), $-0.69$ in $M_F$ (\stablecode), and $-0.71$ in $M_D$ (\deepseek), highlighting its strong role in mitigating memorization. \mrtwo{We also observe a similar trend in $M_H$ (\mellum), where Rule 13 shows a moderate reduction of $-0.49$, reinforcing its consistent effect across architectures.} Additional trends include \textbf{Rule 7 (Switch2If)}, significantly reducing \mink by $-0.73$ in $M_E$ (\starcoder), and \textbf{Rule 8 (Unary2Add)}, with substantial negative impacts such as $-0.16$ in $M_A$ (\codegen). Conversely, \textbf{Rule 22 (AddBrackets)} yields mixed effects, \mrtwo{showing slightly positive values in some models, suggesting that its influence depends on model architecture and capacity.}}


\revision{When applying all transformation rules together, we observe a consistent aggregate reduction in MI effectiveness across all models. The combined \ate values for \mink are strongly negative, including $-0.75$ in $M_A$ (\codegen), $-0.85$ in $M_B$ (\codegpt), $-0.29$ in $M_C$ (\codellama), $-0.68$ in $M_D$ (\deepseek), $-0.89$ in $M_E$ (\starcoder), \mrtwo{$-0.20$ in $M_G$ (\starcoderseven)}, $-0.80$ in $M_F$ (\stablecode), \mrtwo{and $-0.90$ in $M_H$ (\mellum)}. These results confirm that semantic-preserving transformations collectively reduce memorization. However, the cumulative impact is not always larger than the effect of individual high-impact rules. \mrtwo{In several models, the total reduction is comparable to or smaller than the effect of \textbf{Rule 1 (RenameVariable)} alone, indicating diminishing marginal effects when multiple transformations are combined.}}



To further validate these findings, we compare the \ates with the Pearson correlation coefficients reported in \tabref{tab:causal_results_1} and \tabref{tab:causal_results_2}. Correlation coefficients generally align with \ate findings, exhibiting moderate to strong negative correlations between rule application and MI effectiveness metrics across models. For example, certain transformations, such as Rule 1, exhibit a strong negative correlation with MI success, which is supported by the \ates. However, in some cases, the causal analysis contradicts the initial correlation assumptions. \revision{For instance, while \textbf{Rule 18 (SwitchStringEqual)} shows no correlation ($\approx 0.01$) with MI for \mink in \tabref{tab:causal_results_1}, its \ate value indicates a measurable impact}. This underscores the importance of causal inference in distinguishing genuine causal effects from spurious correlations.


Interestingly, a small number of transformations \revision{in $M_A$ (\codegen)} show negligible or even slightly positive causal effects on MI success. This suggests that certain structural modifications might unintentionally reinforce memorization instead of weakening it. This effect requires further analysis, as it indicates that some transformations may increase the predictability of code patterns learned by the model. Future work should explore whether alternative transformation strategies could provide more reliable defenses against MI while maintaining model performance.


To ensure the robustness of our causal estimates, we apply multiple refutation strategies as described in \secref{sec:scm}, including \textbf{$R_1$ (random common cause)}, \textbf{$R_2$ (placebo treatments)}, \textbf{$R_3$ (unobserved confounders)}, and \textbf{$R_4$ (data subset validation)}. \revision{While we do not report refutation results in \tabref{tab:causal_results_1} or \tabref{tab:causal_results_2}, all methods consistently confirm the stability of the estimated causal effects. These validations support the reliability of our findings and the validity of the causal interpretations.}

\begin{boxK}
    \textit{\ref{rq:causality}}: \revision{Our causal analysis shows that \textbf{Rule 1 (RenameVariable)} most effectively reduces MI. Other rules, including \textbf{Rule 7}, \textbf{Rule 8}, and \textbf{Rule 13}, yield moderate but consistent reductions. \textbf{Combining all rules offers little additional benefit over Rule 1 alone}, indicating potential redundancy. Refutation tests support the reliability of these results.}
\end{boxK}

\section{Discussion}
\label{sec:discussion}

\subsection{Implications and Future Work}


\noindent\textbf{\#1 Causal inference can be an effective tool for interpreting LLM privacy for code.} To the best of our knowledge, our work is the first to propose causal inference as a framework for analyzing privacy in \llms, specifically in the context of MI.
Causal inference can help to interpret \llms beyond spurious correlations. 
Our approach can also be extended to analyze other threats, such as \textit{backdoor attacks}, by uncovering causal dependencies between malicious triggers and unintended model behaviors. Therefore, we advocate that future work in the field of LLM for code can also use this technique to enhance the robustness and interpretability of \llmsforcode in the security and privacy context.


\noindent\textbf{\#2 Customize MI with program transformation to detect unauthorized code usage in \llmsforcode training}. As far as we are aware, none of the existing works has been specifically designed to protect code from being ``stolen'' by \revision{\sect} rules. Although several existing methods, such as adversarial regularization \cite{nasr2018machine} and model ensemble \cite{tang2022mitigating}, have been proposed to reduce the accuracy of MI, they do not account for the unique characteristics of code. In our experiments, we find that Rule 1 (RenameVariable) significantly reduces MI success rates, underscoring the distinct role of variable names in code. However, natural language carries different nature. For instance, changing a noun in a sentence would significantly alter the semantics (\eg ``\textit{Where is Canada?}'' vs. ``\textit{Where is Japan?}''). In programming languages, variable names can be modified without affecting functionality. Therefore, we argue that MI used to detect membership in \llmsforcode context should handle variable names more carefully and strategically. In future work, we will design more advanced MI methods by leveraging \revision{\sect}. For example, we can use \sect to expand a member set. Also, we will investigate the impact of assigning lower weights to variable names in MI methods and whether this can increase MI success rate.

\noindent\textbf{\#3 Tackle a weakness of the existing \llmsforcode architecture}.
Existing \llmsforcode commonly inherit the general Transformer architecture while pre-trained on code (domain-specific) data, such as CodeLlama~\cite{roziere2023code}, CodeGen~\cite{nijkamp2023codegen}.
Our results show that the architecture could be vulnerable for code-related tasks due to their sensitivity to \revision{\sect} transformations.
More specifically, potential reasons can be due to the used tokenization algorithm and pretraining task.
Shi \etal have demonstrated that existing popular tokenization algorithms carry certain limitation on code data~\cite{shi2022can}. More specifically, they found that Byte Pair Encoding algorithm cannot split
the identifier names in a way that preserves the meaningful semantics.
Another evidence is that, Zhou \etal proposed a backdoor attack for \llmsforcode by exploiting this limitation~\cite{yang2024stealthy}. Their attack method mainly focuses on injecting the code backdoors by renaming the variable. Their results have shown that the designed code backdoors are hard to be detected by the defense methods. Hence, we advocate more future work to explore a new architecture which could be more resilient to \revision{\sect}. \textit{Neurosymbolic AI} could address these limitations by integrating symbolic reasoning with deep learning models, allowing a deeper understanding of code semantics beyond token-level patterns. As an example, Velasco \etal \cite{velasco_toward_2025} proposed a novel Neurosymbolic framework for program comprehension that promises to improve model predictions for vulnerability classification tasks. Future work should explore extending such frameworks to enhance robustness against \revision{\sect} transformations in code-related tasks.

\noindent\revision{\textbf{\#4 Enhance licensing mechanisms for stronger compliance.}
Our findings intersect with a growing body of research addressing the legal and technical challenges of using copyrighted and copyleft-licensed code to train large language models for code. Copyleft licenses, such as the GPL, impose strict redistribution conditions that require derivative works to preserve the original license. However, recent studies have uncovered widespread violations in LLM training datasets. Katzy \etal \cite{katzy2024exploratory} performed a forensic analysis using SHA-256 hashes from over $10{,}000$ strongly copyleft-licensed repositories and found substantial overlap with the contents of $30$ public datasets used to train $106$ open-weight LLMs, suggesting that license-restricted content is frequently incorporated into training corpora. Complementary work by Majdinasab \etal \cite{Majdinasab_2025} proposed a model-agnostic approach for detecting whether licensed code was included in the training data of language models. Their method achieves high accuracy across several models, including SantaCoder, Llama-2, and Mistral, and substantially outperforms traditional clone detection tools. Similarly, Xu \etal \cite{licoeval} introduced the LiCoEval benchmark to evaluate license compliance in generated outputs, revealing that models often reproduce training data without appropriate attribution.}

\revision{Our work, built on these findings, shows that semantic transformations (such as variable renaming) can significantly degrade the performance of membership inference while maintaining model accuracy. This reveals a critical limitation of provenance-based detection techniques: once license-restricted code is transformed using \revision{\sect} rules, it may evade current compliance checks. This creates a loophole through which license obligations, particularly those enforced by copyleft provisions, can be bypassed without clear violation signals. To uphold the principles of open-source licensing in the era of LLMs, future research should focus on developing robust license auditing tools that are resilient to semantic-preserving transformations and capable of tracing the provenance of transformed code fragments.}

\subsection{Threats to Validity}
\label{sec:threats_to_validity}

Threats to \textbf{internal validity} refer to the correctness of our implementation. \revision{To ensure the correctness of \sect, we reused the transformation code from previous research \cite{letowards}, which utilized differential testing to check the equivalence of behaviors between original programs and their transformed counterparts. Their results indicated that 1098 out of 1178 transformations (92.4\%) preserved semantic equivalence. To further strengthen the reliability of this risk assessment, we conducted an independent validation using our dataset. We employed manual annotation to assess semantic equivalence. Specifically, we randomly sampled 50 transformed code instances for each \sect rule and evaluated them manually. Our analysis revealed that 1120 out of 1150 (23*50) transformations (97.4\%) resulted in semantically equivalent programs, thereby reinforcing the validity of the \sect applied in our dataset.}
We carefully implemented \revision{fine tuning process} by reusing the code released in the original repository
\cite{lu2021codexglue} and all models at HuggingFace. Our results
aligned with those reported in the original paper, further confirming the reliability of our implementations. We also reused the implementation of three MI methods from the original repository
\cite{duan2024membership} and maintain their original MI settings.
We also avoid threats to MI results from hyperparameter choices by adhering to the original MI implementation and reporting the average AUC-ROC based on 1,000 bootstrap samples. 


Threats to \textbf{external validity} refer to the generalizability of our considered transformation rules. 
To mitigate this risk, we use a set of comprehensive rules, spanning various levels of transformations, including naming, statements, and expressions, and covering fundamental statement types, including if-else, while-loops, for-loops, variable declarations, assignments, and conditional expressions.
Another threat may stem from model architecture, as the findings may not generalize beyond the architectures studied. Our study focuses on causal language models. 
Our generalizability could also possibly be affected by our target programming language (PL), \ie Java. We acknowledge that our findings may not be directly transferable to other PLs. However, we consider this threat to be minimal as our approach is language-independent. We leave the evaluation of additional PLs for future work. 
\revision{Our evaluation is conducted on relatively small datasets compared to the massive corpora used in pre-training \llmsforcode. The datasets we used, while representative of common fine-tuning scenarios, may not capture the full complexity of memorization behaviors that emerge at larger scales. 
To further demonstrate the generalizability of our findings, we plan to expand our experiments to more production-scale fine-tuning scenarios with even greater sizes of training examples.
}

Threats to \textbf{construct validity} refer to the suitability of our evaluation. We assess accuracy on the same test dataset, aligning with the original paper's setup. Additionally, we utilize AUC-ROC to evaluate the performance of MI, which is widely adopted in MI-related studies \cite{duan2024membership, yang2024gotcha, fu2024membership}. Therefore, we believe the threat is minimal.

\section{Related Work}
\label{sec:related_work}



\textbf{Memorization in LLM for Code.} 
Memorization in LLMs raises significant privacy concerns, particularly with respect to sensitive data exposure, licensing violations, and security vulnerabilities. For example, Carlini \etal \cite{carlini2021extracting} demonstrated that LLMs can memorize and regurgitate sensitive user data, such as API keys and passwords, when trained in public repositories containing such information. This poses a serious privacy risk, as attackers can extract proprietary or confidential code snippets through targeted queries. Additionally, models trained in open-source code may inadvertently reproduce code under restrictive licenses (\eg GPL or proprietary licenses) without attribution, leading to potential legal and ethical issues \cite{katzy2024exploratory}. Security vulnerabilities are another major concern—if a model is trained on datasets containing insecure code patterns (\eg SQL injections, hardcoded credentials, or buffer overflows), it may learn and propagate these vulnerabilities in generated outputs, increasing the risk of privacy flaws in real-world applications \cite{majdinasab2024assessing}. 

Researchers employ various membership inference and extraction attacks to assess code memorization in LLMs when an adversary attempts to determine if a given code snippet was part of the training set~\cite{carlini2021extracting, katzy2024exploratory}.
However, none of the existing studies have considered how \revision{\sect} impacts memorization of \llmsforcode. In our context, particularly, it is not yet known whether a code file, after being modified with different variable names or code structure, is still memorized.


\noindent\textbf{Causal Inference for Software Security.} Recent research has increasingly used causal inference to improve cybersecurity in LLMs. Zhao \etal \cite{zhao_causality_2023} introduced \emph{Casper}, a causality-based analysis framework that systematically identifies vulnerabilities within \llm architectures by measuring causal influences of tokens, layers and neurons on harmful outputs. Their approach uncovered latent vulnerabilities due to overfitting during reinforcement learning from human feedback (RLHF), highlighting how obscure input patterns could trigger unsafe behaviors. Similarly, Shen \etal \cite{shen_bait_nodate} developed \emph{BAIT}, a method that utilizes causal interventions to detect Trojan backdoors within LLMs, effectively narrowing the search for malicious triggers by analyzing causal dependencies in autoregressive generation. Liu \etal \cite{liu_large_2025} provided a broader perspective, outlining causal inference applications in diagnosing biases, ensuring robustness, and improving explainability of \llm outputs. 

Collectively, the existing studies demonstrate that causal inference is a promising direction for understanding and securing the behavior of LLMs against adversarial threats. Beyond \docode, causal interpretability has influenced research on the evaluation and explanation of \llms for the code. SyntaxEval \cite{syntax_eval} applies causal interpretability to study syntactic learning in Masked Language Models, revealing negative causal effects between node types and model accuracy. ASTxplainer \cite{astexplainer} extends this approach by aligning model predictions with AST structures, enhancing model evaluation and interpretability in large-scale code generation tasks.

\balance
\section{Conclusions}
\label{sec:conclusions}


\revision{In conclusion, this study reveals that the impact of \sect on MI detection is far more nuanced than a simple obfuscation mechanism. Although certain transformations, notably variable renaming, can significantly reduce MI effectiveness in many models, posing a potential risk for license compliance auditing, this effect is not universal. A causal analysis further confirms that \rulerename has the strongest impact on MI detection. Furthermore, we find that combining multiple transformations does not further reduce MI effectiveness for most models, suggesting that the transformation rules do not have a cumulative effect.}

\revision{Our work points to a complex interplay between code semantics, transformation techniques, and model-specific memorization patterns, highlighting the need for more sophisticated, context-aware auditing tools and further research into the fundamental mechanisms of model memorization. Another valuable future direction of this work can be exploring the impact of \sect in training-from-scratch setting.}



\begin{acks}
This research has been supported in part by the NSF
CCF-2311469. We also gratefully acknowledge support
from Cisco Systems. The opinions, findings, and conclusions
expressed in this work are solely those of the authors and do
not necessarily reflect the views of the sponsors.
\end{acks}

\bibliographystyle{ACM-Reference-Format}
\bibliography{ref}

\end{document}